%% file: ParallelSFL.tex
\documentclass[sigconf,10pt]{acmart}
\usepackage{pifont}
\usepackage{balance}
\usepackage{enumitem}
\usepackage{graphicx}
\usepackage{subfigure}
\usepackage{multirow}
\usepackage{tablefootnote}
\usepackage{threeparttable}
\usepackage[ruled,vlined,linesnumbered,ruled]{algorithm2e}
\usepackage{color}
\usepackage{algorithm2e}
\usepackage[noend]{algpseudocode}
\usepackage{xspace}
\usepackage{fancyhdr}

\usepackage{tikz}
\newcommand*{\circled}[1]{\lower.8ex\hbox{\tikz\draw (0pt, 0pt)%
    circle (.47em) node {\makebox[0.4em][c]{\small #1}};}}

\def\ie{\textit{i.e.}\xspace}

\def\etal{\textit{et al.}\xspace}

\def\eg{\textit{e.g.}\xspace}

\def\method{ParallelSFL\xspace}

\newcommand{\bluenote}[1]{\textcolor{black}{#1}}

\AtBeginDocument{%
  \providecommand\BibTeX{{%
    \normalfont B\kern-0.5em{\scshape i\kern-0.25em b}\kern-0.8em\TeX}}}

\begin{document}

\setcopyright{acmlicensed}
\copyrightyear{2024}
\acmYear{2024}

\acmConference[ACM MobiCom '24]{International Conference On Mobile Computing And Networking}{Nov. 18-22, 2024}{Washington, DC, USA}
\acmBooktitle{International Conference On Mobile Computing And Networking (ACM MobiCom '24), November 18-22, 2024, Washington D.C., DC, USA}
\acmDOI{10.1145/3636534.3690665}
\acmISBN{979-8-4007-0489-5/24/09}

\title{\textit{\method}: A Novel Split Federated Learning Framework Tackling Heterogeneity Issues}


\author{Yunming Liao$^{1,2}$, *Yang Xu$^{1,2}$, *Hongli Xu$^{1,2}$, Zhiwei Yao$^{1,2}$\\Liusheng Huang$^{1,2}$, Chunming Qiao$^{3}$}
\affiliation{%
  \institution{$^{1}$School of Computer Science and Technology, University of Science and Technology of China\\
    $^{2}$Suzhou Institute for Advanced Research, University of Science and Technology of China\\
    $^{3}$Department of Computer Science and Engineering, University at Buffalo, the State University of New York\\
    \{ymliao98, zhiweiyao\}@mail.ustc.edu.cn,~\{xuyangcs, xuhongli, lshuang\}@ustc.edu.cn,~qiao@buffalo.edu}
\city{}
\country{}
\thanks{* Corresponding authors.}
}

\renewcommand{\shortauthors}{\footnotesize Yunming Liao, Yang Xu, Hongli Xu, Zhiwei Yao, Liusheng Huang, Chunming Qiao}



\begin{abstract}
\input{contents/abstract.tex}

\end{abstract}


	


\maketitle


\section{Introduction}\label{sec:intro}
\input{contents/intro.tex}

\section{Related Work}\label{sec:relwork}
\input{contents/works.tex}

\vspace{-0.05cm}
\section{Preliminary}\label{sec:prelim}
\input{contents/motivations.tex}


\section{System Design}\label{sec:design}
\input{contents/design.tex}

\section{Evaluation}\label{sec:evaluation}
\input{contents/experiment.tex}

\section{Conclusion}\label{sec:conclusion}
\input{contents/conclusion.tex}

\section*{Acknowledgment}
This article is supported in part by the National Key Research and Development Program of China (No. 2021YFB3301500); in part by the National Science Foundation of China (NSFC) under Grants 62102391, 61936015, and 62132019; in part by USTC Research Funds of the Double First-Class Initiative (No. WK2150110030).

\balance
\bibliographystyle{unsrt}
\bibliography{contents/refs}

\end{document}

%% file: contents/abstract.tex
Mobile devices contribute more than half of the world’s web traffic, providing massive and diverse data for powering various federated learning (FL) applications.
In order to avoid the communication bottleneck on the parameter server (PS) and accelerate the training of large-scale models on resource-constraint workers in edge computing (EC) system, we propose a novel split federated learning (SFL) framework, termed \textit{\method}.
Concretely, we split an entire model into a bottom submodel and a top submodel, and divide participating workers into multiple clusters, each of which collaboratively performs the SFL training procedure and exchanges entire models with the PS.
However, considering the statistical and system heterogeneity in edge systems, it is challenging to arrange suitable workers to specific clusters for efficient model training.
To address these challenges, we carefully develop an effective clustering strategy by optimizing a utility function related to training efficiency and model accuracy.
Specifically, \method partitions workers into different clusters under the heterogeneity restrictions, thereby promoting model accuracy as well as training efficiency.
Meanwhile, \method assigns diverse and appropriate local updating frequencies for each cluster to further address system heterogeneity.
Extensive experiments are conducted on a physical platform with 80 NVIDIA Jetson devices, and the experimental results show that \method can reduce the traffic consumption by at least 21\%, speed up the model training by at least 1.36$\times$, 
and improve model accuracy by at least 5\% in heterogeneous scenarios, compared to the baselines.

%% file: contents/intro.tex

With the development of Smart Home, various privacy data (\eg, monitoring data, sensor data) are collected by home devices and stored in the smart gateways without sharing with others \cite{stojkoska2017review, bansal2020iot}.
However, these privacy data are extremely valuable and urgently needed for training to further improve the performance of on-device intelligent applications, such as fall detection \cite{yacchirema2018fall}, health monitoring \cite{chaudhary2021multi, sathasivam2022iot}, smart fire detection and surveillance \cite{al2019smart, komalapati2021smart}, intrusion detection \cite{elrawy2018intrusion}, \bluenote{and natural language processing \cite{vaswani2017attention} } for Smart Home.
To extract knowledge in the vast amount of privacy data, \textit{edge AI} has been proposed and become a dominant tool \cite{rudman2016defining,wang2019edge,li2019edge}.
One of the popular techniques in edge AI is federated learning (FL), which trains a globally-shared model through collaboration among smart gateways in the data-parallel fashion \cite{hard2018federated, kairouz2019advances, han2021legodnn, mcmahan2017communication}.

\begin{figure}[!t]
	\centering
	\includegraphics[width=1\columnwidth]{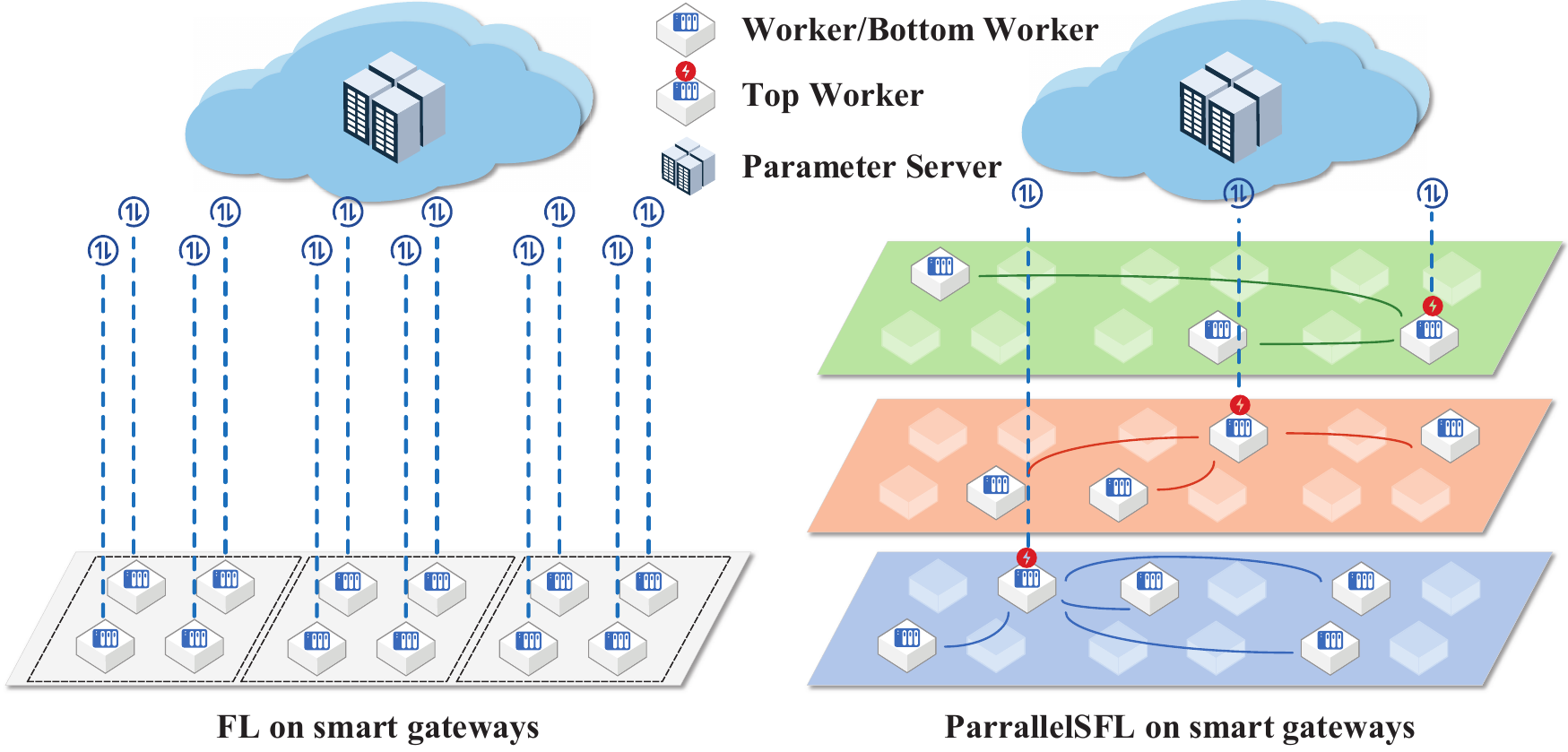}
	\caption{Illustration of FL and \method.}
	\label{fig:scenario}
\end{figure}

In FL, each worker (\eg, a smart gateway) is responsible for training an entire model and periodically pushes/pulls the updated/aggregated model to/from the parameter server (PS) until model convergence, as illustrated in Fig. \ref{fig:scenario}.
To boost the performance of AI applications/services, it is necessary and practical to augment the parameters of deep learning models, \bluenote{\eg, large language models (LLMs) with transformer architectures \cite{liu2019roberta, nakkiran2021deep, niu2020billion, zhang2021elf}.}
However, due to the hardware limitations of the resource-constrained workers, which usually are only equipped with 1$\sim$30TOPS computing power and 1$\sim$8GB memory, the requirements of high computing power and large memory for model training hinder each worker from training a complete large-scale model \cite{park2021communication, pal2021server, liao2023mergesfl, oh2022locfedmix}.

Hence, split federated learning (SFL) has been proposed by incorporating both data parallelism and model parallelism to train large-scale models \bluenote{such as large CNN models and LLMs} \cite{pal2021server, thapa2022splitfed, han2021accelerating, abedi2020fedsl}.
SFL splits an entire model into two submodels, \ie, bottom submodel and top submodel, at the split layer.
The bottom submodel is trained on the resource-constrained workers, while the top submodel is offloaded to the PS. 
Then the workers perform the training procedure by continuously exchanging the smashed data \bluenote{(also called activations)} and gradients with the PS.
However, considering the contradiction between the limited bandwidth of PS and numerous workers, the PS may become the system bottleneck, leading to the risk of network congestion and poor scalability \cite{kairouz2019advances}.

To avoid the communication bottleneck of the PS, we propose a novel SFL framework, termed \textit{\method}, which partitions workers into multiple clusters, and encourages each cluster to collaboratively perform the SFL training procedure and exchange entire models with the PS, as illustrated in Fig. \ref{fig:scenario}.
Within each cluster, one worker (denoted as the top worker) maintains the top submodel while the remaining workers (denoted as bottom workers) train the bottom submodels.
They together perform local updating by exchanging smashed data and gradients.
After local updating, the top worker aggregates bottom submodels, and sends bottom and top submodels to the PS for splicing and aggregation.
With the benefits of split learning, \method can not only release the computing burden on workers for training large-scale models, but also avoid the communication bottleneck and reduce the network traffic of the PS.


However, apart from the resource limitation, \method still suffers from two other critical challenges in practical applications.
1) \textbf{\textit{Statistical Heterogeneity.}} 
Since the local data collected by workers depend on their user preferences and are not shared with others for privacy concerns in Web 3.0, resulting in non-independent and identically distributed (non-IID) data across all workers \ie, statistical heterogeneity \cite{zhao2022fedgan, zhuang2022divergence, liao2023adaptive, wang2022accelerating, xie2023federatedscope}.
The non-IID data slows down the convergence rate and even compromises the accuracy of the trained models \cite{zhao2018federated, wang2020optimizing, li2022pyramidfl}.
2) \textbf{\textit{System Heterogeneity.}} Workers generally are configured with varying and limited capabilities in EC systems.
The computing and communication capabilities of workers could differ from each other by more than tenfold times \cite{chen2022decentralized, lai2021oort, xu2022adaptive}.
System heterogeneity poses significant influences on synchronous training processes, as fast workers may be forced to wait for slow ones, leading to increased waiting time and decreased training efficiency \cite{lim2020federated, zhang2018adaptive, li2022pyramidfl}.

To this end, \method is designed to simultaneously tackle the heterogeneity issues through careful and effective worker clustering based on the distinct properties of SFL, which is fundamentally different from the clustering strategy in existing FL works (more details in Sec. \ref{sec:design}).
On one hand, to address system heterogeneity,
it is necessary to shrink the waiting time across workers in each cluster as far as possible, by organizing workers with similar computing/communication capabilities into the same cluster.
On the other hand, motivated by the existing works \cite{mhaisen2021optimal}, the local data of each cluster together (except the top worker) should be close to IID to deal with statistical heterogeneity.
In \method, we introduce the KL-divergence to measure the gap between the data distribution of each cluster and IID, and optimize the gap to enhance model accuracy.
However, it is intricate to develop the clustering strategy under the heterogeneity restrictions, so as to balance the trade-off between training efficiency and model accuracy.
Meanwhile, we assign diverse and appropriate local updating frequencies for clusters to further reduce the waiting time across clusters.
In a nutshell, our main contributions are as follows:
\begin{itemize}
    \item We propose a novel SFL framework, named \method, which is designed to overcome the resource limitation and address system as well as statistical heterogeneity by effective cluster partitioning with the distinct properties of SFL.
    \item We define a utility function with the heterogeneity restrictions to serve as a comprehensive metric for estimating waiting time and data distribution of worker clusters. Upon this, we develop the clustering strategy to balance the trade-off between training efficiency and model accuracy.
    \item We evaluate the performance of \method through a physical platform with totally 80 NVIDIA Jetson edge devices.
    The experimental results show that the \method can reduce the traffic consumption by at least 21\%, speed up the model training by at least 1.36$\times$, and improve model accuracy by at least 5\% in heterogeneous scenarios, compared to the baselines.
\end{itemize}

The rest of the paper is organized as follows.
Sec. \ref{sec:relwork} reviews some related works of SFL and FL.
Sec. \ref{sec:prelim} introduces the background of FL as well as SFL and present our proposed SFL framework.
Sec. \ref{sec:design} elaborates the detailed design of our framework.
In Sec. \ref{sec:evaluation}, we perform extensive experiments to evaluate our framework.
Finally, the whole paper is concluded in Sec. \ref{sec:conclusion}.

%% file: contents/works.tex
\subsection{Split Federated Learning}
\bluenote{The previous vanilla SL methods \cite{gupta2018distributed, vepakomma2018split} are proposed to help workers collaboratively train DL models without sharing sensitive raw data in domains such as health care and finance.
At that time, the server needs to communicate with workers one by one to complete model training, leading to poor flexibility and scalability.}
By incorporating FL with SL, SplitFed \cite{thapa2022splitfed} firstly demonstrates the feasibility and superiority of SFL, and aggregates bottom models after each local updating.
Such frequent aggregation results in high network traffic consumption.
To save the traffic consumption, LocSplitFed \cite{han2021accelerating} allows the workers not to send features to the PS by using local-loss-based training, which cannot update the top model in time and results in much computing resource.
Then, LocFedMix-SL \cite{oh2022locfedmix} is implemented to maintain all the benefits of SplitFed and LocSplitFed with fixed local updating frequency, but still cannot fully utilize the capacities of heterogeneous workers.
\bluenote{In addition, Kim \etal \cite{kim2023bargaining} and Joshi \etal \cite{joshi2021splitfed} propose SFL approaches without aggregating bottom models to reduce the traffic consumption and accelerate model training, but sacrificing model accuracy.}
The existing SFL works mainly focus on training large-scale DL models on resource-constrained workers.
Although Liao \etal \cite{liao2023accelerating} present an advanced solution (\ie, AdaSFL), which assigns adaptive and diverse batch sizes for different workers to address system heterogeneity, AdaSFL still cannot deal with the statistical heterogeneity.
Despite these notable advancements, none of the existing SFL works have yet explored to simultaneously address system and statistical heterogeneity.

\subsection{Federated Learning}
Prior to the emergence of SFL, many solutions to address the heterogeneity challenges \cite{liao2023decentralized, luo2022tackling,arisdakessian2023towards,li2022pyramidfl,lai2021oort, xu2022adaptive} have been studied in typical FL scenarios.
In order to alleviate the negative effect of system heterogeneity, Diao \etal \cite{diao2020heterofl} propose HeteroFL, which enables the training of heterogeneous local models with varying computation complexities on different workers.
Besides, Xu \etal \cite{ xu2022adaptive} investigate to optimize the local updating frequency of different workers, where the workers with high computing/communication capacities are assigned with larger local updating frequencies.
\bluenote{To deal with statistical heterogeneity, Sattler \etal \cite{sattler2020clustered} propose IFCA, the clustered FL method to alternate between estimating the cluster identities and minimizing the loss functions}, while Shin \etal \cite{shin2022fedbalancer} propose FedBalancer to actively select clients' important training samples.
In addition, other works \cite{luo2022tackling,li2022pyramidfl,lai2021oort} propose to employ worker selection to simultaneously address system and statistical heterogeneity. 
Specifically, Li \etal \cite{li2022pyramidfl} develop PyramidFL, a fine-grained worker selection strategy that focuses on the divergence between the selected workers and the remaining workers to fully exploit the computing resource and data of different workers.
In addition, Luo \etal \cite{luo2022tackling} propose AdaSampling and design an adaptive worker sampling algorithm, which tackles both system and statistical heterogeneity to minimize the wall-clock training time.
There are also many existing works \cite{ghosh2020efficient, vahidian2023efficient, ghosh2022efficient} on clustering strategies in FL to deal with system and statistical heterogeneity.
However, those FL researches still face the challenge of training large-scale models on resource-constrained workers.
Furthermore, those FL researches can not be directly applied for SFL and \method, since workers maintaining only the bottom models must exchange smashed data/gradients with the PS (or the top worker) keeping the top model continuously.

%% file: contents/motivations.tex
\begin{figure*}[t]
    \centering
    \includegraphics[width=1\linewidth]{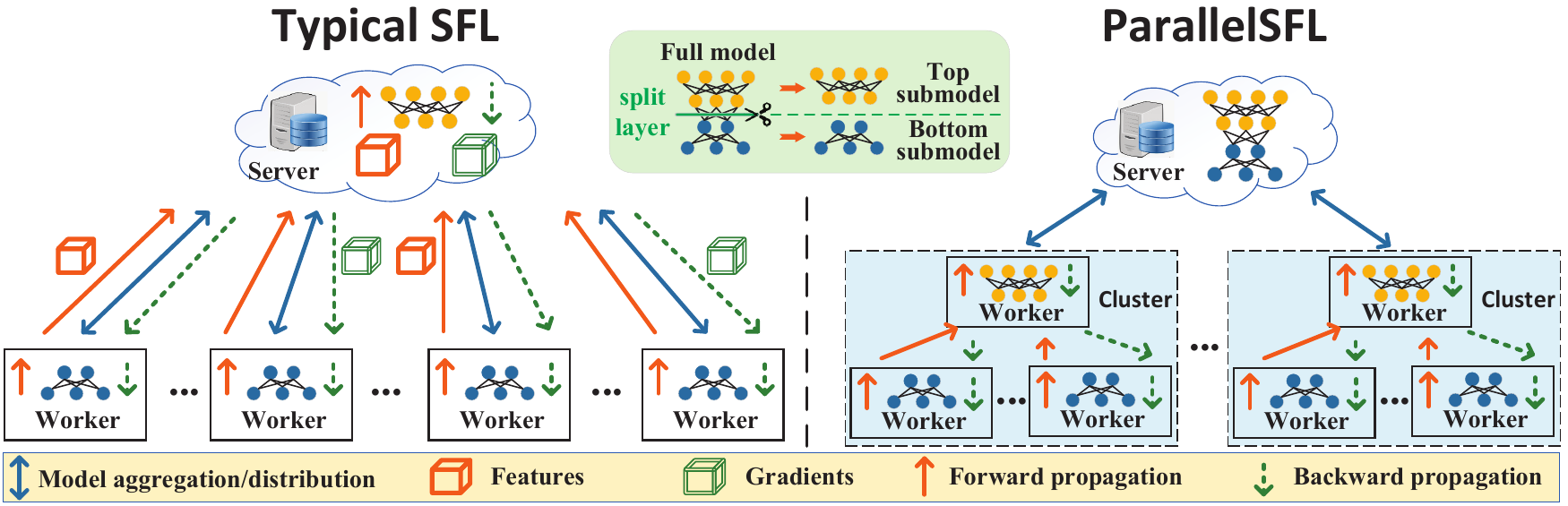}
    \caption{Illustration of typical SFL and \method.}
    \label{fig:comparison}
\end{figure*}
    
\subsection{Federated Learning}
There are $N$ workers and a parameter server (PS) constituting an EC system, where FL is implemented to perform the learning tasks through a decentralized collaboration among the participating workers.
In FL, the PS maintains a globally shared model $\boldsymbol{w}$ with complete structure, while each worker $i$ ($\in [N]$) holds its local data $\mathbb{D}_i$ and trains a local model $\boldsymbol{w}_i$ (\ie, a replica of the global model $\boldsymbol{w}$).
The objective of FL is to find the optimal model $\boldsymbol{w}^*$ that minimizes the loss function, which can be defined as:
\begin{equation}\label{equ:opti}
    \min _{\boldsymbol{w}} F(\boldsymbol{w}) \triangleq \frac{1}{N} \sum_{i=1}^{N} F_i(\boldsymbol{w}_i)
\end{equation}
where $F_i(\boldsymbol{w}_i)=\frac{1}{|\mathbb{D}_{i}|} \sum_{x \in \mathbb{D}_{i}} \ell\left(x ; \boldsymbol{w}_{i}\right)$ is the local loss function of worker $i$, and $\ell\left(x ; \boldsymbol{w}_{i}\right)$ is the loss with respect to the model $\boldsymbol{w}_i$ and the data sample $x$.

Due to the inherent complexity of most DL tasks, it is both essential and functional to employ the gradient descent algorithms \cite{wang2018edge, ma2021adaptive} to solve Eq. (\ref{equ:opti}).
In each aggregation round, worker $i$ downloads the global model $\boldsymbol{w}$ from the PS and independently trains local model $\boldsymbol{w}_i$ using its own data $\mathbb{D}_{i}$.
The PS aggregates the models from all workers and obtains the updated global model $\boldsymbol{w}=\frac{1}{N}\sum_{i=1}^{N} \boldsymbol{w}_i$ and moves on to the next round.
In doing so, the global model can acquire knowledge from the local data of different workers without leaking their data privacy.
However, training large-scale models in typical FL remains challenging, as it imposes a significant computation burden on resource-constrained workers and entails substantial communication overhead for model exchange between the PS and workers \cite{han2021accelerating,oh2022locfedmix}.

\subsection{Split Federated Learning}
SFL serves as an alternative solution to process complex real-world data using a large-scale model.
The fundamental concept of SFL is to split the full model $\boldsymbol{w}$ into two submodels at the split layer, denoted as $\boldsymbol{w}=[\boldsymbol{w}_b, \boldsymbol{w}_p]$, where $\boldsymbol{w}_b$ represents the bottom submodel and $\boldsymbol{w}_p$ represents the top submodel.
In the case of a CNN model, the bottom submodel typically comprises the input layer and convolutional layers whereas the top submodel consists of fully-connected layers and the output layer.
In SFL, the PS holds the top submodel $\boldsymbol{w}_p$, while each worker $i$ trains a bottom submodel $\boldsymbol{w}_{b,i}$ using its local data $\mathbb{D}_i$.
By training only the submodels, the computation burden of workers can be significantly reduced, compared to FL

As illustrated in Fig. \ref{fig:comparison} (left plot), the basic training process of SFL involves three main stages, \ie, forward/backward propagation of the worker-specific bottom submodels, forward/backward propagation of the top submodel, and global aggregation of bottom submodels on the PS.
Firstly, each worker performs forward propagation with a batch of data samples, and delivers the smashed data of the split layer to the PS.
Subsequently, the PS conducts forward/backward propagation to update the top submodel.
Then, the PS sends the corresponding gradients back to the workers, enabling them to update their bottom submodels through backward propagation.
After local updating, the PS aggregates the bottom submodels from all workers and sends the aggregated bottom submodels back to the workers for further training.
However, SFL requires continuous exchange of smashed data/gradients between workers and the PS, which consumes the available bandwidth of the PS and generates a substantial amount of traffic workload.
Consequently, the PS may become a system bottleneck, leading to the risk of network congestion and poor scalability.

\subsection{Our Proposed SFL Framework}
To alleviate the computing/communication burden on the resource-constrained workers and the PS, we propose a novel SFL framework, called \textit{\method}, which is designed to tackle heterogeneity issues.
Fig. \ref{fig:comparison} (right plot) illustrates the \method framework, where we partition the $N$ workers into $C^h$ clusters in aggregation round $h$.
In the cluster $c$, there are $N_c$ bottom workers training the bottom submodels and one designated top worker keeping the top submodel.
Within each cluster, the bottom workers and the top worker collaborate to perform local updating by exchanging smashed data and gradients.
After several local iterations, the top worker aggregates bottom submodels, and sends the bottom submodel and top submodel to the PS for splicing and aggregation.
In round $h$, the bottom submodel on worker $i$ at iteration $k$ is denoted as $\boldsymbol{w}_{b,i}^{h,k}$, and one iteration for updating the bottom submodel is expressed as:
\begin{equation}
    \boldsymbol{w}_{b,i}^{h,k+1}=\boldsymbol{w}_{b,i}^{h,k}- \eta \widetilde{\nabla} F_{b,i}(\boldsymbol{w}_{b,i}^{h,k})
\end{equation}
Here, $\widetilde{\nabla} F_{b,i}(\boldsymbol{w}_{b,i}^{h,k})=\frac{1}{|D_{i}|} \sum_{x \in D_{i}} \nabla \ell(x ; \boldsymbol{w}_{b,i}^{h,k})$ is the gradient for a certain mini-batch $D_{i}$, and $\nabla \ell(x ; \boldsymbol{w}_{b,i}^{h,k})$ denotes the stochastic gradient given the bottom submodel $\boldsymbol{w}_{b,i}^{h,k}$ and input data sample $x$.
In round $h$, let $\boldsymbol{w}_{p,c}^{h,k}$ represent the top submodel on the top worker of cluster $c$ at the iteration $k$.
Then, the process of one iteration on the top worker is expressed as:
\vspace{-0.05cm}
\begin{equation}\label{server-update}
    \boldsymbol{w}_{p,c}^{h,k+1}=\boldsymbol{w}_{p,c}^{k} - \frac{\eta}{N_c} \sum_{i=1}^{N_c} \widetilde{\nabla} F_{p,i}(\boldsymbol{w}_{p,c}^{h,k})
    \vspace{-0.05cm}
\end{equation}
where $\widetilde{\nabla} F_{p,i}(\boldsymbol{w}_{p,c}^{h,k})=\frac{1}{|D_{i}|} \sum_{x \in D_{i}} \nabla \ell(\boldsymbol{w}_{b,i}^{h,k}(x); \boldsymbol{w}_{p,c}^{h,k})$ is the gradient of the top submodel's loss function, and $\nabla \ell(\boldsymbol{w}_{b,i}^{h,k}(x); \boldsymbol{w}_{p,c}^{h,k})$ denotes the stochastic gradient for the top submodel $\boldsymbol{w}_{p,c}^{h,k}$ and the output of the bottom submodel $\boldsymbol{w}_{b,i}^{h,k}$ (\ie, smashed data), given the input data sample $x$.
After local updating, the top worker aggregates the bottom submodels from all bottom workers in cluster $c$ as follows:
\vspace{-0.05cm}
\begin{equation}\label{eq:bottom_aggregation}
    \boldsymbol{w}_{b,c}^{h}=\frac{1}{N_c}\sum_{i=1}^{N_c} \boldsymbol{w}_{b,i}^{h}
    \vspace{-0.05cm}
\end{equation}
Subsequently, the top worker in each cluster sends the aggregated bottom submodel and the top submodel to the PS.
Finally, the PS splices the submodels into the entire model $w_c^h=[w_{b,c}^h, w_{p,c}^h]$ corresponding to cluster $c$ and aggregates them to obtain the updated global model as follows:
\vspace{-0.1cm}
\begin{equation}\label{global_aggregation}
    \boldsymbol{w}^{h+1}= \frac{1}{C^h} \sum_{c=1}^{C^h} w_c^h
    \vspace{-0.1cm}
\end{equation}
After that, the PS distributes the updated global model $\boldsymbol{w}^{h+1}$ to the top worker in new clusters and moves on to the next aggregation round.

To tackle the heterogeneity issues and avoid the communication bottleneck, it is crucial to carefully group the workers into proper clusters to enhance model accuracy and improve training efficiency simultaneously.
In Sec. \ref{sec:partitioning_strategy}, we develop the clustering strategy under the heterogeneity restrictions.
Although we organize workers with similar computing/communication capabilities into the same cluster, the complete time of one local iteration among clusters still varies significantly considering the heterogeneous workers in diverse clusters.
In traditional synchronous schemes, if we assign identical local updating frequencies for all clusters, the fast clusters are forced to wait for the slow ones, incurring idle waiting time and significantly destroying the training efficiency.
Generally, the clusters with higher performance can be allocated with larger local updating frequencies, while those with lower performance are allocated with smaller ones.
Therefore, the waiting time among clusters would be greatly reduced.

%% file: contents/design.tex
\begin{figure}[!t]
	\centering
	\includegraphics[width=1\columnwidth]{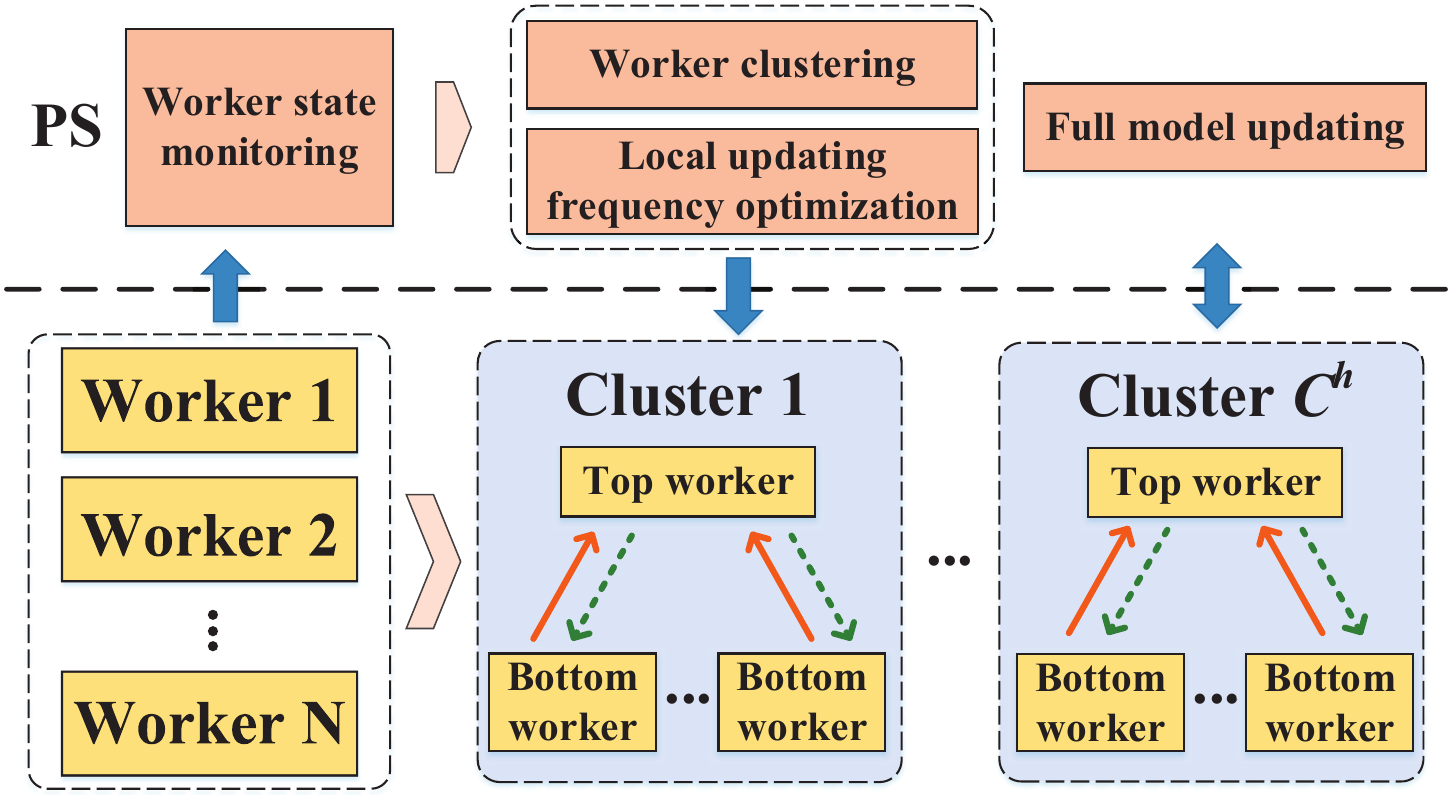}
	\caption{Overview of \method.}
	\label{fig:overview}
    \vspace{-0.3cm}
\end{figure}

\subsection{Overview} \label{sec:overview}
The overall design of \method is illustrated in Fig. \ref{fig:overview}.
\method consists of four key modules at the PS side, \ie, \textit{worker state monitoring}, \textit{worker clustering}, \textit{local updating frequency optimization}, and \textit{full model updating}.
The worker side mainly consists of the local updating module.
At the beginning of each aggregation round, the worker state monitoring module starts to collect the state information (\eg, ingress bandwidth, label distribution, computing and communication capabilities) from all workers.
After that, the worker clustering module in the PS organizes the workers into suitable clusters to tackle heterogeneity issues.
Subsequently, the local updating frequency optimization module determines diverse and appropriate local updating frequencies for clusters to further reduce the waiting time across clusters.
Meanwhile, the full model updating module distributes the full model to the top worker of each cluster, and then the top worker delivers the bottom submodels to the bottom workers in the cluster.
Within each cluster, the bottom workers continuously exchange smashed data/gradients with the top worker at each iteration.
After a certain number of local iterations, the top worker aggregates the bottom submodels from bottom workers in each cluster.
Then, the PS splices and aggregates the top and bottom submodels from the top workers in each cluster, and moves on to the next aggregation round.

\subsection{Worker State Monitoring} \label{sec:worker_state_collecting}
In order to make an effective clustering strategy, it is necessary to monitor the current states of all workers (\eg, ingress bandwidth, label distribution, computing and communication capabilities).

In EC, each worker $i$ has a limited available ingress bandwidth $B_i^h$ in each round $h$.
The top worker in each cluster typically consumes a significant portion of the bandwidth to exchange smashed data/gradients with bottom workers.
To prevent the top worker from becoming a bottleneck, it is instinct to ensure that the occupied bandwidth of the top worker does not exceed the available ingress bandwidth.
At the beginning of a certain round, the PS collects and analyzes the statistical distribution of the ingress bandwidth of all workers.
Then the PS employs the statistical results as the available ingress bandwidth for each worker.

Secondly, the local data of each cluster together (except the top worker) should be close to IID.
Herein, the label distribution, a vector $\textbf{V}=\{v_j \geq 0, j\in [1,M]\}$ ($\sum_{j=1}^{M}v_j=1\}$) to parameterize a categorical distribution of class labels over $M$ classes, is required to assist in worker clustering.
In typical SFL, the workers deliver the smashed data with corresponding labels to continue forward propagation of the top submodel on the PS, which enables the PS to directly collect the labels of workers' smashed data and generates the label distribution $\textbf{V}_i$ of the mini-batch of worker $i$ \cite{thapa2022splitfed, han2021accelerating, pal2021server}.

Thirdly, the collection of time-varying computing and communication capacities of workers is necessary for worker clustering and local updating frequency optimization.
On one hand, we denote the computing time of one iteration on the bottom submodel and top submodel for worker $i$ in round $h$ as $\mu_{b,i}^h$ and $\mu_{p,i}^h$, respectively.
The relationship between these two parameters is clear and fixed, so that we only monitor $\mu_{b,i}^h$.
On the other hand, $\beta_{i,j}^h$ is indicated as the transmission time of the smashed data/gradients at one iteration between worker $i$ and worker $j$ in round $h$.
As proxy metrics, we adopt $\mu_{b,i}^h$ and $\beta_{i,j}^h$, which can be recorded by the workers directly during model training, to indicate the computing and communication capacities of worker $i$ in round $h$, respectively.
Prior to starting model training in round $h$, the PS collects the latest $\hat{\mu}_{b,i}^h$ and $\hat{\beta}_{i,j}^h$ from all workers.
Besides, to improve the robustness of the estimation, we introduce the moving average with the historical states of workers \cite{leroy2019federated}.
Accordingly, the PS estimates $\mu_{b,i}^{h}$ and $\beta_{i,j}^{h}$ on worker $i$ in aggregation round $h$ by calculating the moving average with $\alpha \in [0, 1]$ (\eg, 0.8 in our experiments) as:
\begin{equation}
\mu_{b,i}^{h}=\alpha \cdot \mu_{b,i}^{h-1}+ (1-\alpha) \cdot \hat{\mu}_{b,i}^h
\end{equation}
\begin{equation}
\beta_{i,j}^{h}=\alpha \cdot \beta_{i,j}^{h-1}+ (1-\alpha) \cdot \hat{\beta}_{i,j}^h
\end{equation}

Since the size of the information about worker states (\eg, 100-300KB \cite{lyu2018multi}) is much smaller than that of model parameters, it is reasonable to ignore the cost (\eg, bandwidth consumption and time cost) for state monitoring \cite{lyu2019optimal}.

\subsection{Worker Clustering} \label{sec:partitioning_strategy}
Based on the collected state information, \method partitions the workers into $C^h$ clusters and determines the top worker for each cluster.
Besides, each cluster is configured with suitable local updating frequency.

In each cluster $c$, there are $N_c$ bottom workers and one designated top worker.
Due to the constraint of the available ingress bandwidth $B_c^h$ for the top worker, the number of participating workers $N_c$ to simultaneously train the bottom submodels is limited.
We denote the bandwidth occupied by each worker at each iteration as $b$.
The occupied bandwidth in each cluster is limited as follows:
\begin{equation}\label{eq:bandwidth_constraint}
    N_c \cdot b \le B_c^h
\end{equation}
The size of cluster $c$ depends on the available ingress bandwidth $B_c^h$ of the top worker.
Since the local data of the top worker does not contribute to model training, we should control the number of the top worker (\ie, the number of the clusters) as small as possible.

\begin{algorithm}[t] \caption{Worker clustering in round $h$} \label{alg}
\KwIn{$B_i^h$, $\textbf{V}_i$, $\mu_{b,i}^h$, $\mu_{p,i}^h$, $\beta_{i,j}^h$, $K$.}
\KwOut {The worker clustering strategy.}

Calculate $KL(\Phi_i || \Phi_j)$ for all workers $i$ and $j$, $i \neq j$.\\

Divide the workers into $K$ sets $S_1, \dots, S_K$ ($S = S_1 \cup \dots \cup S_K$) by calling the $K$-means algorithm. \\

Initialize $c=1$. \\

\While{$S$ is not empty}{
Select the worker $i$ with maximum $B_i^h$ from $S_k$ with the most workers as the top worker for cluster $c$. \\

Remove worker $i$ from its set. \\

Denote the distribution of cluster $c$ as $\Phi_c$. \\

Select the worker $j$ with the maximum $t_j^h$ from each $S_k$ into set $A$. \\

Select workers into the cluster $c$ from set $A$ to make $KL(\Phi_0 || \Phi_c)$ smallest under the constraints of Eqs. \eqref{eq:bandwidth_constraint} and \eqref{eq:top_worker_constraint}. \\
$c \leftarrow c+1$.\\
}

Minimize the utility function $\sum_{c=1}^{C^h} \mathcal{U}_c$ by exchanging workers from different clusters, without violating Eqs. \eqref{eq:bandwidth_constraint} and \eqref{eq:top_worker_constraint}.
\end{algorithm}

For the cluster $c$, we can formulate the completion time $t_i^h$ of one iteration on the worker $i$ in round $h$ as:
\begin{equation}
    t_i^h=\mu_{b,i}^h+\beta_{i,c}^h+\mu_{p,c}^h
\end{equation}
where $\mu_{p,c}^h$ and $\beta_{i,c}^h$ separately denote the computing time of one iteration of the top worker and the transmission time of the smashed data at one iteration from bottom worker $i$ to the top worker in cluster $c$.
Since the top worker has to simultaneously exchange smashed data/gradients with multiple workers, in order to ensure training efficiency, the number of participating workers $N_c$ in cluster $c$ is limited as:
\begin{equation}\label{eq:top_worker_constraint}
    N_c \cdot \mu_{p,c}^h \leq \text{max}\{\mu_{b,i}^h+\beta_{i,c}^h\}, \ \forall i \in [N_c]
\end{equation}
Besides, the waiting time of the bottom worker $i$ can be defined as $t_{c,o}^h-t_i^h$, where $t_{c,o}^h=\text{max}\{t_i^h\}$ ($\forall i \in [N_c]$) denotes the completion time of one iteration in round $h$ for the slowest worker in cluster $c$.
Accordingly, the average waiting time of cluster $c$ at one iteration in round $h$ is formulated as:
\begin{equation}
    \mathcal{W}_c^h =\frac{1}{N_c} \sum_{i=1}^{N_c}(t_{c,o}^h-t_i^h)
\end{equation}
To minimize the waiting time of all workers in cluster $c$, we should group the $N_c$ workers, whose completion time of one iteration is close enough to each other, into the same cluster.

To tackle the non-IID issue, the data distribution of each cluster needs to be close to IID.
We first define the IID distribution as $\Phi_0$.
If the data of all workers follows IID distribution, we get $\Phi_0=\frac{1}{N}\sum_{i=1}^N \textbf{V}_i$, where $\textbf{V}_i$ is the label distribution of worker $i$.
In cluster $c$, the label distribution of $N_c$ workers is:
\begin{equation}
    \Phi_c = \frac{1}{N_c}\sum_{i=1}^{N_c}\textbf{V}_i
\end{equation}
Then, we introduce the KL-divergence $KL(\Phi_c || \Phi_0)$ to measure the gap between $\Phi_c$ and $\Phi_0$ as follows \cite{hershey2007approximating, goldberger2003efficient}:
\begin{equation}\label{eq:KL}
    KL(\Phi_c || \Phi_0) = \sum_{j=1}^{M} \Phi_c(v_j) \text{log} \frac{\Phi_c(v_j)}{\Phi_0(v_j)}
\end{equation}
To deal with statistical heterogeneity, it is necessary to control the KL-divergence $KL(\Phi_c || \Phi_0)$ as small as possible.

Taking system and statistical heterogeneity into account, it is challenging to partition these workers into appropriate clusters.
We normalize $\mathcal{W}_c^h$ as well as $KL(\Phi_c || \Phi_0)$, and introduce a utility function to evaluate the effect of cluster $c$ in round $h$ as follows:
\begin{equation}
    \mathcal{U}_c = \lambda  \cdot \mathcal{W}_c^h + (1-\lambda) \cdot KL(\Phi_c || \Phi_0)
\end{equation}
where $\lambda$ is a weight coefficient used to balance $\mathcal{W}_c^h$ and $KL(\Phi_c || \Phi_0)$.
In round $h$, we need to carefully partition all the workers into appropriate clusters to minimize $\sum_{c=1}^{C^h} \mathcal{U}_c$ under the constraints of Eqs. \eqref{eq:bandwidth_constraint} and \eqref{eq:top_worker_constraint}, so that we can simultaneously address system and statistical heterogeneity and implement efficient \method.

We propose a greedy algorithm to make the effective clustering strategy.
Firstly, by the $K$-means algorithm (\eg, $K=N/5$), according to the KL-divergence of label distribution among workers, we divide the workers with small KL-divergence into the same set and obtain $K$ sets $S_1, \dots, S_K$ (Line 1-2 of Alg. \ref{alg}).
Next, we greedily select the workers with maximum ingress bandwidth $B_c^h$ from the set $S_k$ with the most workers as the top worker for cluster $c$.
We construct the set $A$ including the worker $j$ with the maximum $t_j^h$ from each set $S_k$ and group workers into the cluster $c$ from set $A$ to make $KL(\Phi_0 || \Phi_c)$ smallest under the constraints of Eqs. \eqref{eq:bandwidth_constraint} and \eqref{eq:top_worker_constraint}.
Subsequently, we repeat the above operations to further partition the remaining workers and create new clusters, until all workers are grouped into suitable clusters (Line 4-10 of Alg. \ref{alg}).
Finally, we optimize the distribution of workers among clusters through fine-tuning, aiming to minimize the utility function $\sum_{c=1}^{C^h} \mathcal{U}_c$, while adhering to the constraints specified in Eqs. \eqref{eq:bandwidth_constraint} and \eqref{eq:top_worker_constraint} (Line 11 of Alg. \ref{alg}).
Upon completion of this process, we obtain the $C^h$ effective clusters in round $h$.



\subsection{Local Updating Frequency Optimization} \label{sec:local_updating_frequency_configuration}
Additionally, due to the system heterogeneity, the complete time of one iteration among clusters is highly different.
We denote the communication time for the top worker in cluster $c$ transmitting the submodels to the PS in round $h$ as $\beta_c^h$.
The local updating frequency of cluster $c$ is defined as $\tau_c^h$.
Accordingly, the completion time $t_c^h$ of aggregation round $h$ for the cluster $c$ is expressed as:
\begin{equation}
    t_c^h=\tau_c^h \cdot t_{c,o}^h + \beta_c^h
\end{equation}
Besides, the waiting time of cluster $c$ can be defined as $t^h-t_c^h$, where $t^h=\text{max}\{t_c^h\}$ ($\forall c \in [C^h]$) denotes the completion time of aggregation round $h$ for the slowest cluster.
Accordingly, the average waiting time of all clusters in round $h$ is formulated as:
\begin{equation}
    \mathcal{W}^h =\frac{1}{C^h} \sum_{c=1}^{C^h}(t^h-t_c^h)
\end{equation}
To minimize the average waiting time, \method regulates the local updating frequencies of all clusters so as to align their completion time of one round. 
Thus, it ensures that the average waiting time will be small enough to mitigate the negative impacts of the synchronization barrier and improve training efficiency.
The regulation rule is expressed as:
\begin{equation}\label{eq:waiting_constraint}
    \lfloor \frac{\tau_c^h \cdot t_{c,o}^h + \beta_c^h}{\tau_l^h \cdot t_{l,o}^h + \beta_l^h} \rfloor = 1 
\end{equation}
where $l$ denotes the index of the fastest clusters assigned with the default maximum local updating frequency $\tau$ in round $h$.
According to Eq. \eqref{eq:waiting_constraint}, we can obtain the specific local updating frequencies for all clusters in round $h$

\subsection{Full Model Updating}\label{full_model_aggregation/distribution}
After cluster $c$ performs local updating in round $h$, the top worker firstly aggregates the bottom submodels from other workers according to Eq. \eqref{eq:bottom_aggregation}.
Then the top worker splices the bottom submodel and top submodel, and sends the full model to the PS for global aggregation.
Considering the clusters with diverse numbers of workers are configured with different local updating frequencies for model training, the full model in each cluster is updated and trained with varying degrees, which needs adaptive aggregation weight to guarantee the performance of aggregated global model \cite{xu2022adaptive, li2022auto, li2023revisiting}.
Therefore, the PS aggregates the entire models from different clusters with adaptive weights related to the local updating frequencies and the size of clusters as follows:
\begin{equation}
\boldsymbol{w}^{h}=\sum_{c=1}^{C^h}\frac{N_c \cdot \tau_c^h \cdot \boldsymbol{w}_{c}^{h}}{\sum_{c=1}^{C^h} N_c \cdot \tau_c^h}
\end{equation}
The aggregated global model is stored in the PS and will be distributed to future clusters to continue further training, or be used for AI tasks.

\subsection{Discussion of Privacy Protection}
\bluenote{In existing SFL literature \cite{thapa2022splitfed, gupta2018distributed, vepakomma2018split}, there are two ways, \ie, sharing labels and returning logits (without sharing labels), to complete the model updating, which can also be applied in ParallelSFL.
Generally, the bottom workers usually deliver the features with corresponding labels to continue forward propagation of the top model on the top worker.
However, if the bottom workers are unwilling or not permitted to share labels, the top worker can return the output logits of the top model to the workers for calculating the loss locally.
Then the bottom workers send the loss values back to the top worker for deriving the gradients and complete the backward propagation.
Moreover, the existing privacy-preserving techniques such as Differential Privacy \cite{thapa2021advancements, wu2023federated, ghazi2021deep} and Homomorphic Encryption \cite{yang2023dynamic} can be employed to further protect the privacy of the smashed data in ParallelSFL.
In this paper, we focus on efficient model training on resource-constrained workers in case of system and statistical heterogeneities, which is orthogonal to privacy protection.}

%% file: contents/experiment.tex
\subsection{Experimental Settings}
\textbf{System Deployment.}
We conduct extensive experiments to evaluate the performance of \method on an edge computing hardware prototype system.
Specifically, we employ a deep learning GPU workstation as the PS, which is equipped with an Intel(R) Core(TM) i9-10900X CPU, four NVIDIA GeForce RTX 2080Ti GPUs and 256 GB RAM.
In addition, we specify 80 NVIDIA Jetson kits\footnote{\url{https://docs.nvidia.com/jetson/}}, including 30 Jetson TX2 devices, 40 Jetson NX devices, and 10 Jetson AGX devices, as workers to construct a heterogeneous system. 
The detailed technical specifications of Jetson TX2, NX, and AGX are listed in Table \ref{table:jetson}.
Notably, the TX2 showcases a 256-core Pascal GPU with 6 TOPs and a CPU cluster consisting of a 2-core Denver2 and a 4-core ARM CortexA57.
The NX is outfitted with a 384-core NVIDIA Volta GPU with 21 TOPs and a 6-core NVIDIA Carmel ARMv8.2 CPU.
Jetson Xavier NX dramatically enhances the NVIDIA software stack over 10$\times$ the performance of Jetson TX2.
Lastly, the AGX stands out with a 512-core NVIDIA Volta GPU with 32 TOPs and an 8-core NVIDIA Carmel ARMv8.2 CPU.
Besides, the computing power of these devices is at the same level as the current mainstream smart gateways.

\begin{table}[!t]
\caption{Device technical specifications.}
\label{table:jetson}
\centering
\begin{tabular}{lcc}
\hline
    & \textbf{AI Performance} & \textbf{GPU Type}  \\ 
\hline
Jetson TX2 & 6 TOPs & 256-core Pascal \\ 
Jetson NX  & 21 TOPs & 384-core Volta\\ 
Jetson AGX  & 32 TOPs & 512-core Volta\\ \hline \hline
& \textbf{CPU Type} & \textbf{ROM} \\  \hline
Jetson TX2 & Denver 2 and ARM 4 & 8 GB LPDDR4\\ 
Jetson NX & 6-core Carmel ARM 8 & 8 GB LPDDR4x\\ 
Jetson AGX  & 8-core Carmel ARM 8 & 32 GB LPDDR4x \\  
\hline \hline
& \textbf{CPU Frequency} & \textbf{GPU Frequency} \\  \hline
Jetson TX2 & 2.0GHz & 1.12GHz\\ \hline
Jetson NX & 1.9GHz & 1.1GHz\\ \hline
Jetson AGX & 2.2GHz & 1.37GHz\\ \hline
\end{tabular}
\end{table}

In the experiments, we build the software platform based on Docker Swarm \cite{merkel2014docker,naik2016building} and the PyTorch deep learning library \cite{paszke2019pytorch}.
The Docker Swarm, a distributed software development kit, facilitates the construction of a distributed system and enables the monitoring of each device's operational status.
The PyTorch library facilitates the implementation of model training on devices. 
Additionally, to streamline communication among devices, we implement MPI (Message Passing Interface) \cite{gabriel2004open}, which includes a collection of sending and receiving functions. 

\textbf{Settings of System Heterogeneity.}
To enable the workers with heterogeneous computing and communication capabilities, we present the following experimental settings.

1) \textbf{\textit{For Computation.}}
All the Jetson TX2, NX, and AGX devices can be configured to work with different modes, which specifies the number of working CPUs and the frequency of CPU/GPU for the devices to work with different computing capacities.
Specifically, TX2 can work in four modes each while NX and AGX work in one of eight modes.
Devices working in different modes exhibit diverse capabilities.
For instance, the AGX with the highest performance mode (\ie, mode 0 of AGX) achieves training by 100$\times$ faster than the TX2 with the lowest performance mode (\ie, mode 1 of TX2).
To further reflect the time-varying on-device resource, we randomly change the modes for devices every 20 rounds.

2) \textbf{\textit{For Communication.}}
All devices are connected to the PS via WiFi routers.
We arrange 80 devices into four groups, each containing 20 devices.
These groups are then placed at different locations, \ie, 2m, 8m, 14m, and 20m away from the WiFi routers. 
Due to random channel noise and competition among devices, the bandwidth between the PS and devices dynamically varies during the training.
The bandwidth of devices is measured by iperf3 \cite{tirumala1999iperf}, which dynamically fluctuates between 1Mb/s and 30Mb/s during training.

\textbf{Applications and Models.}
We evaluate the performance of \method on four applications (\ie, classical datasets) with four DNN models.


1) \textbf{\textit{Speech Recognition.}}
The Google Speech dataset \cite{warden2018speech} (expressed as Speech for short) is adopted for the task of speech recognition, which allows a computer or device to recognize and interpret spoken language.
The dataset includes 85,511 and 4,890 audio clips for training and test, respectively.
The model trained on Speech is a plain CNN model with four 1-D convolutional layers and one fully-connected layer
\footnote{\url{https://pytorch.org/tutorials/intermediate/speech_command_classification_with_torchaudio_tutorial.html}}.

2) \textbf{\textit{Object Recognition.}}
We adopt the CIFAR-10 dataset \cite{krizhevsky2010convolutional} for the evaluation, which is an image dataset composed of 60,000 32$\times$32 color images (50,000 for training and 10000 for test) across 10 categories.
We utilize an 8-layer AlexNet with a size of 136MB \cite{krizhevsky2012imagenet} for CIFAR-10.
The AlexNet is composed of three 3$\times$3 convolutional layers, one 7$\times$7 convolutional layer, one 11$\times$11 convolutional layer, two fully-connected hidden layers, and one softmax output layer.

3) \textbf{\textit{Image Classification.}} 
ImageNet \cite{russakovsky2015imagenet} is a dataset for image recognition that consists of 1,281,167 training images, 50,000 validation images, and 100,000 test images from 1000 categories.
To adapt to the resource-constrained workers, we create a subset of ImageNet, called IMAGE-100, which contains 100 out of 1,000 categories.
We adopt a famous large model VGG16 with a size of 321MB \cite{simonyan2014very} for the complex task, which consists of 13 convolutional layers with the kernel of 3$\times$3, two fully-connected layers, and a softmax output layer.

4) \textbf{\textit{Natural Language Processing.}}
\bluenote{We utilize the Question-answering Natural Language Inference (QNLI) dataset \cite{wang2018glue}, a classification dataset consisting of question-sentence pairs, in this application.
There are 104,743 samples for training and 5,463 samples for test in the QNLI dataset.}
We train a RoBERTa model with a size of 501MB \cite{liu2019roberta} on QNLI dataset.
The RoBERTa is composed of 12 transformer layers, a 768$\times$2 fully-connected layer, and a softmax output layer.

\textbf{Settings of Statistical Heterogeneity.}
In the experiments, training samples of each worker are drawn independently by a vector $\textbf{v}$.
To create non-IID datasets, we draw from a Dirichlet distribution \cite{hsu2019measuring, yurochkin2019bayesian}, \ie, $\textbf{v}\sim\textit{Dir}(\delta \textbf{q})$, where $\textbf{q}$ characterizes a prior class distribution, and $\delta>0$ is a concentration parameter controlling the identicalness among workers.
With $\delta \rightarrow \infty$, all workers have identical distributions to the prior class distribution (\ie, IID); with $\delta \rightarrow 0$, each worker holds data samples from one class, which indicates a high degree of statistical heterogeneity.
We specify 6 values (\eg, $\infty$, 1, 0.5, 0.25, 0.2, 0.1) for $\delta$ to generate different data distributions that cover a spectrum of identicalness, and define $p=1/{\delta}$ (\ie, $p=0,1,2,4,5,10$) to quantify the non-IID levels.
The degree of statistical heterogeneity increases as $p$ increases, and $p$ = 0 is a special case of IIDness.

\begin{figure*}[t]
\centering
\subfigure[Speech]
{
    \includegraphics[width=0.225\linewidth,height=3.3cm]{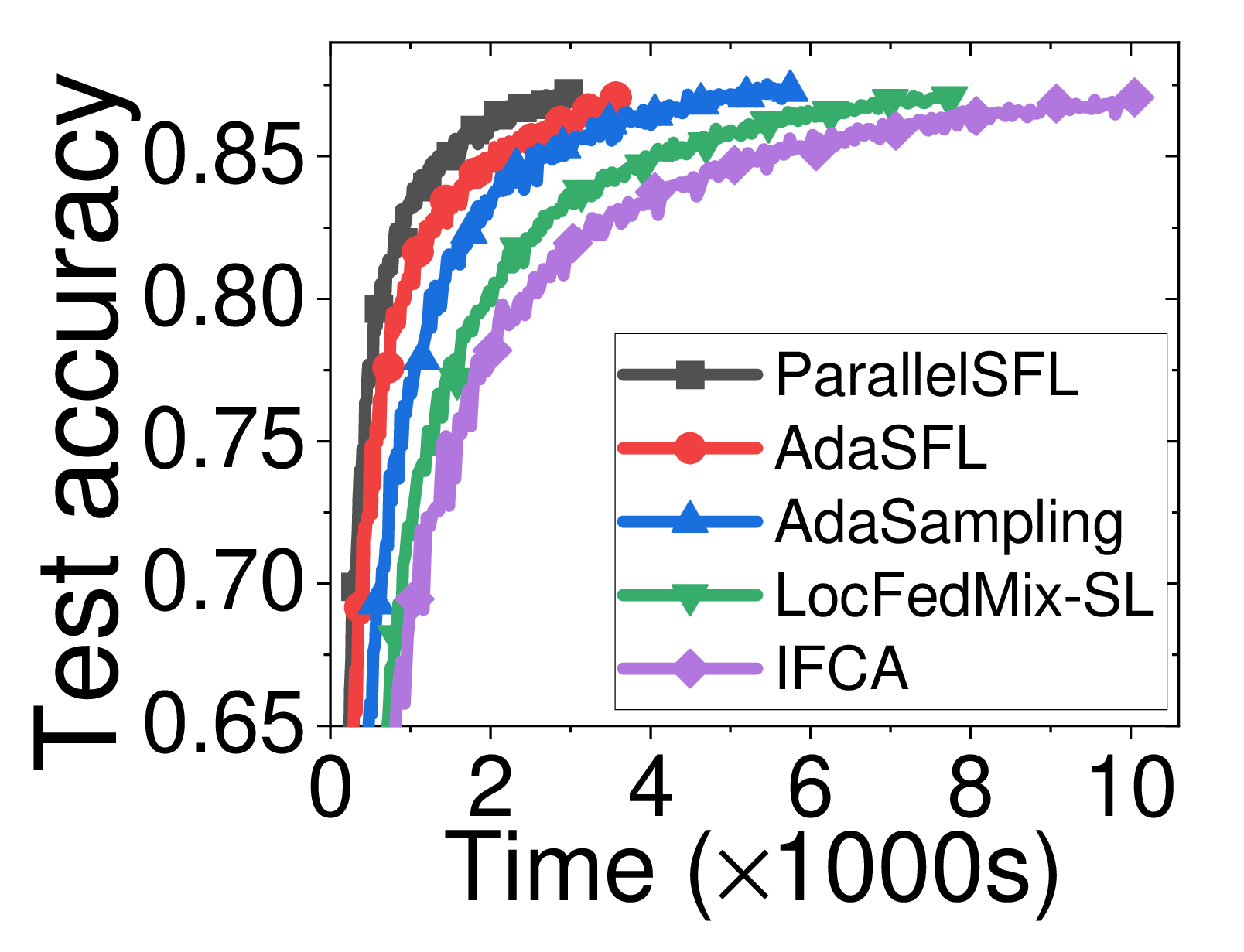}
    \label{fig:acc_iid_speech}
}\quad 
\subfigure[CIFAR-10]
{
    \includegraphics[width=0.225\linewidth,height=3.3cm]{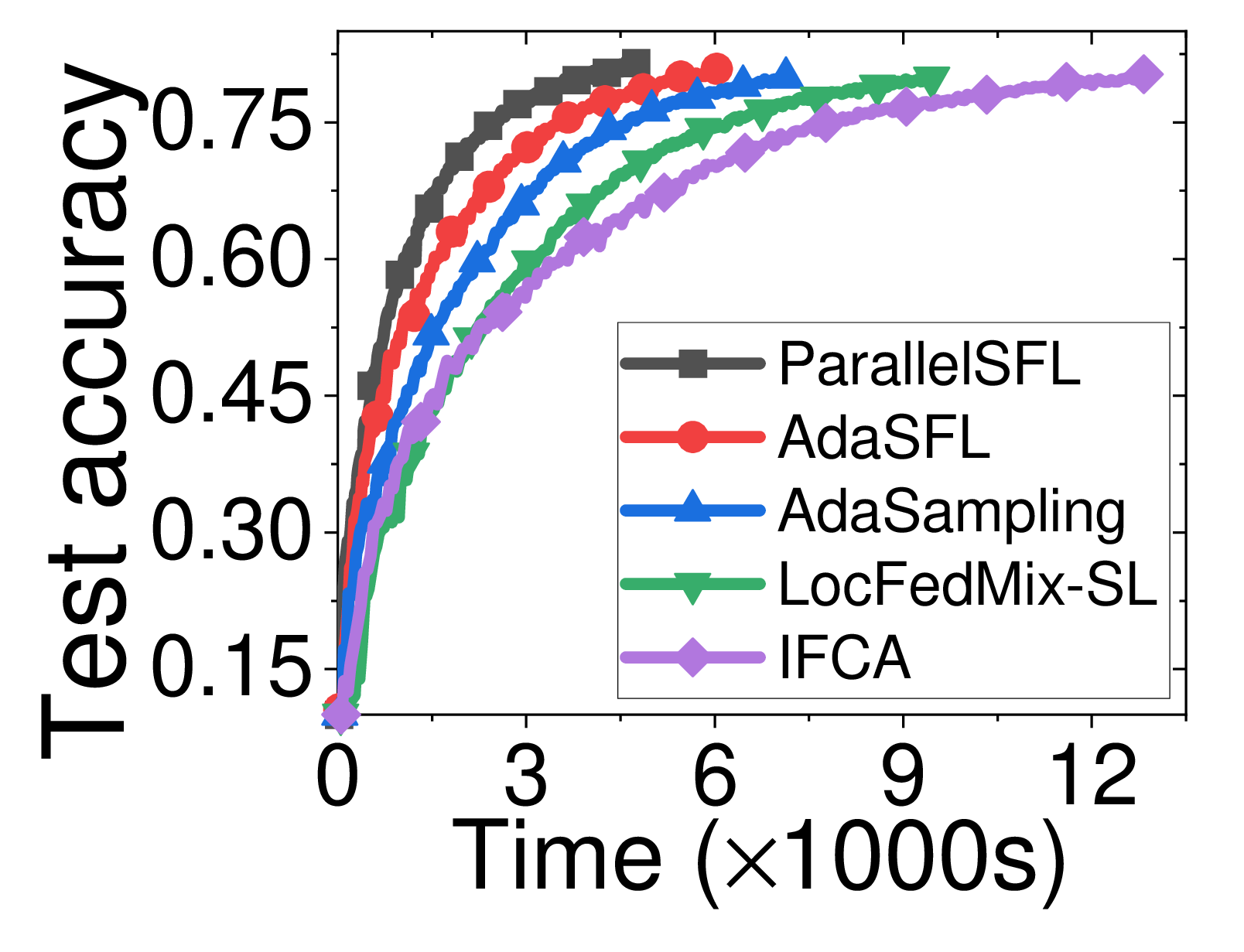}
    \label{fig:acc_iid_CIFAR10}
}\quad 
\subfigure[IMAGE-100]
{
    \includegraphics[width=0.225\linewidth,height=3.3cm]{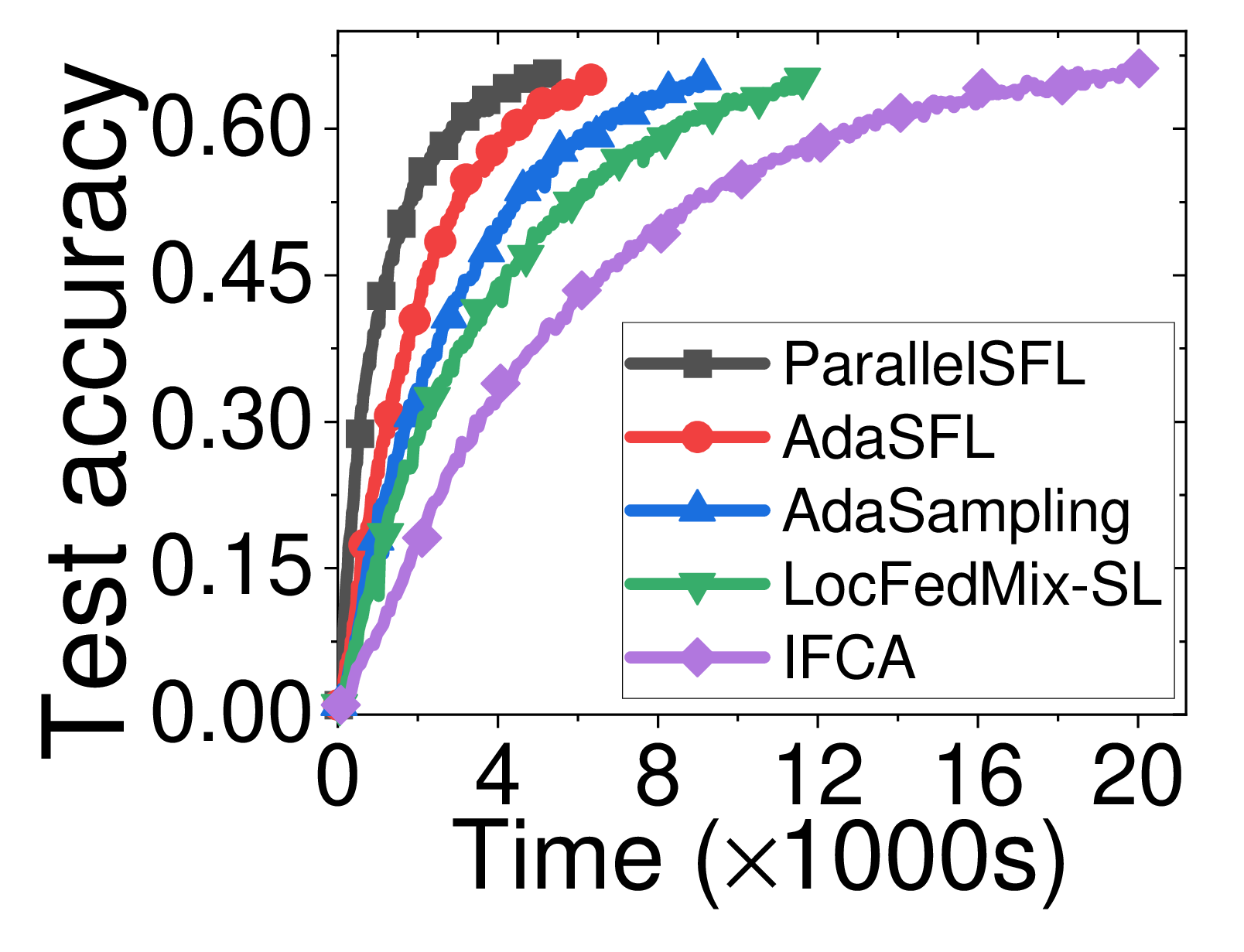}
    \label{fig:acc_iid_image}
}\quad
\subfigure[QNLI]
{
    \includegraphics[width=0.225\linewidth,height=3.3cm]{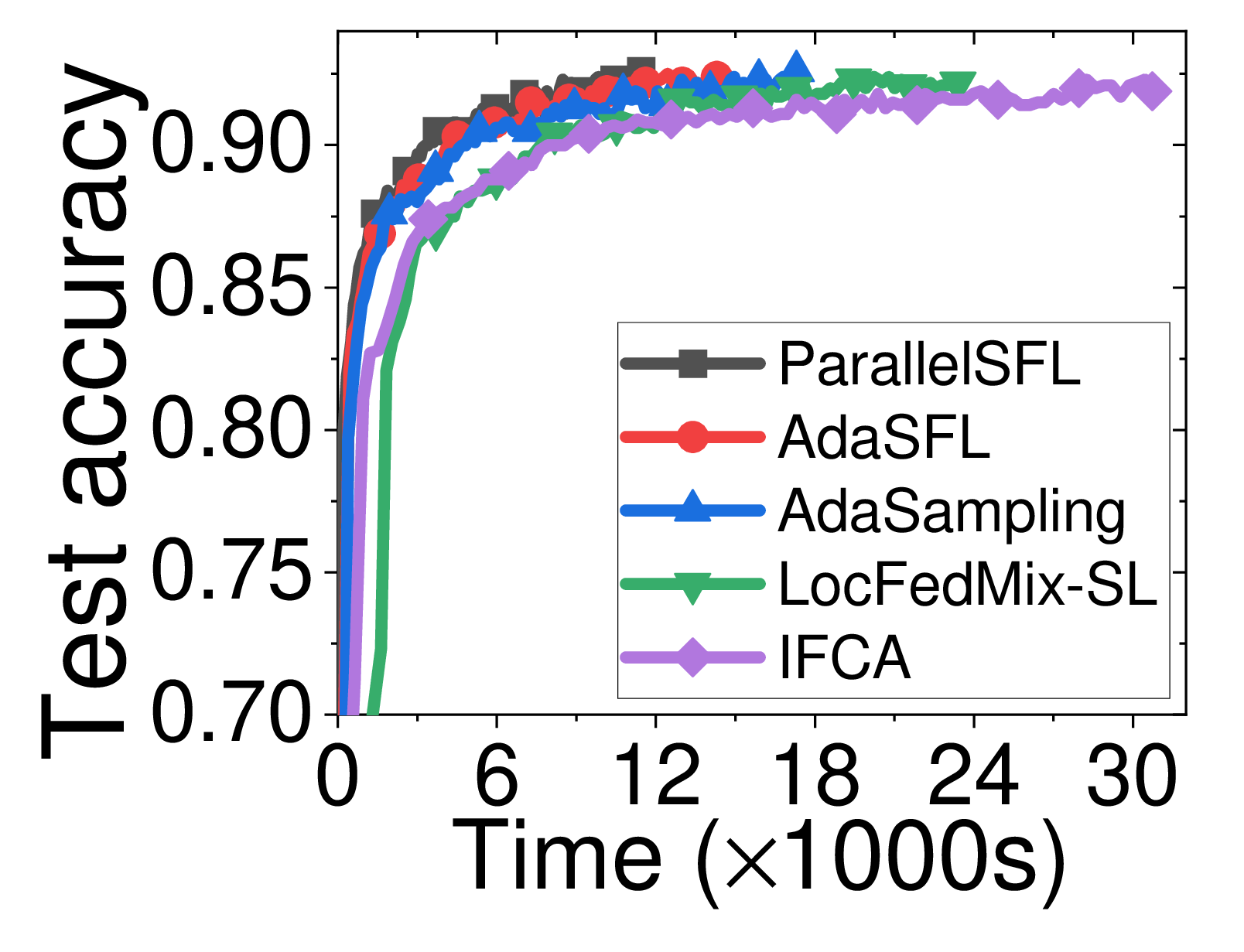}
    \label{fig:acc_iid_Bert}
}
\caption{Test accuracy of five approaches on the four IID datasets.}
\label{fig:acc_iid}
\end{figure*}

\begin{figure*}[!t]
\centering
\subfigure[Speech]
{
    \includegraphics[width=0.225\linewidth,height=3.3cm]{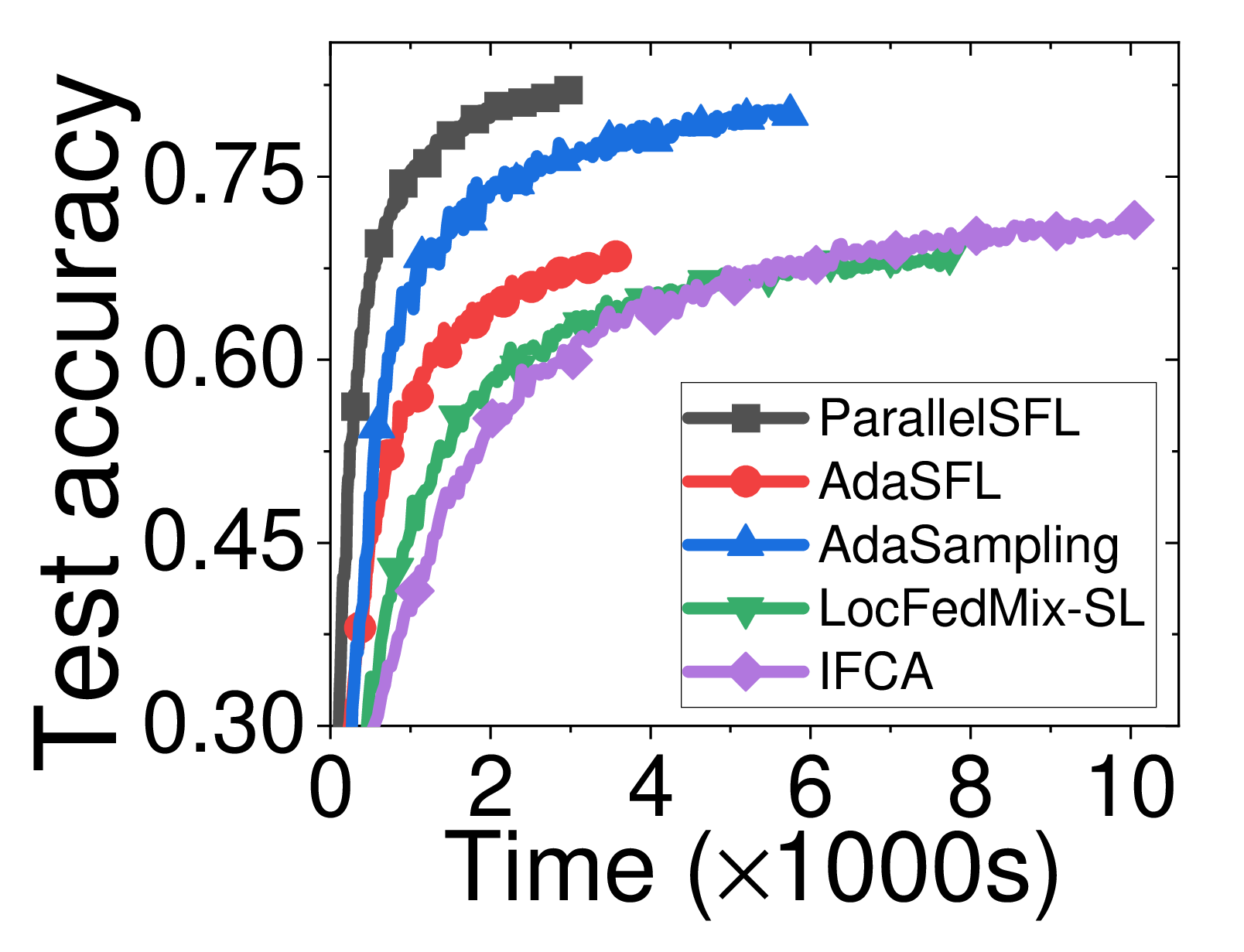}
    \label{fig:acc_noniid_speech}
}\quad 
\subfigure[CIFAR-10]
{
    \includegraphics[width=0.225\linewidth,height=3.3cm]{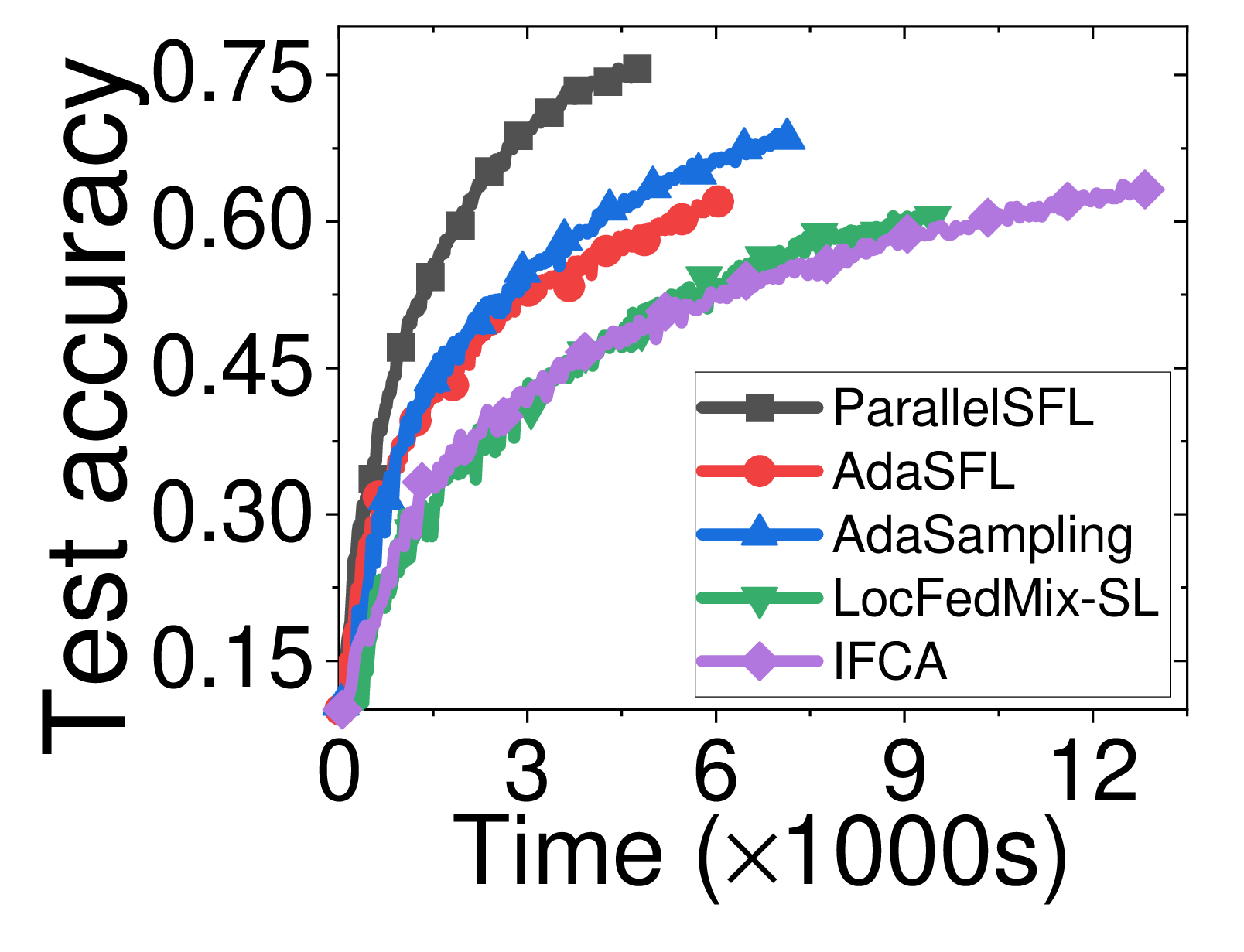}
    \label{fig:acc_noniid_CIFAR10}
}\quad 
\subfigure[IMAGE-100]
{
    \includegraphics[width=0.225\linewidth,height=3.3cm]{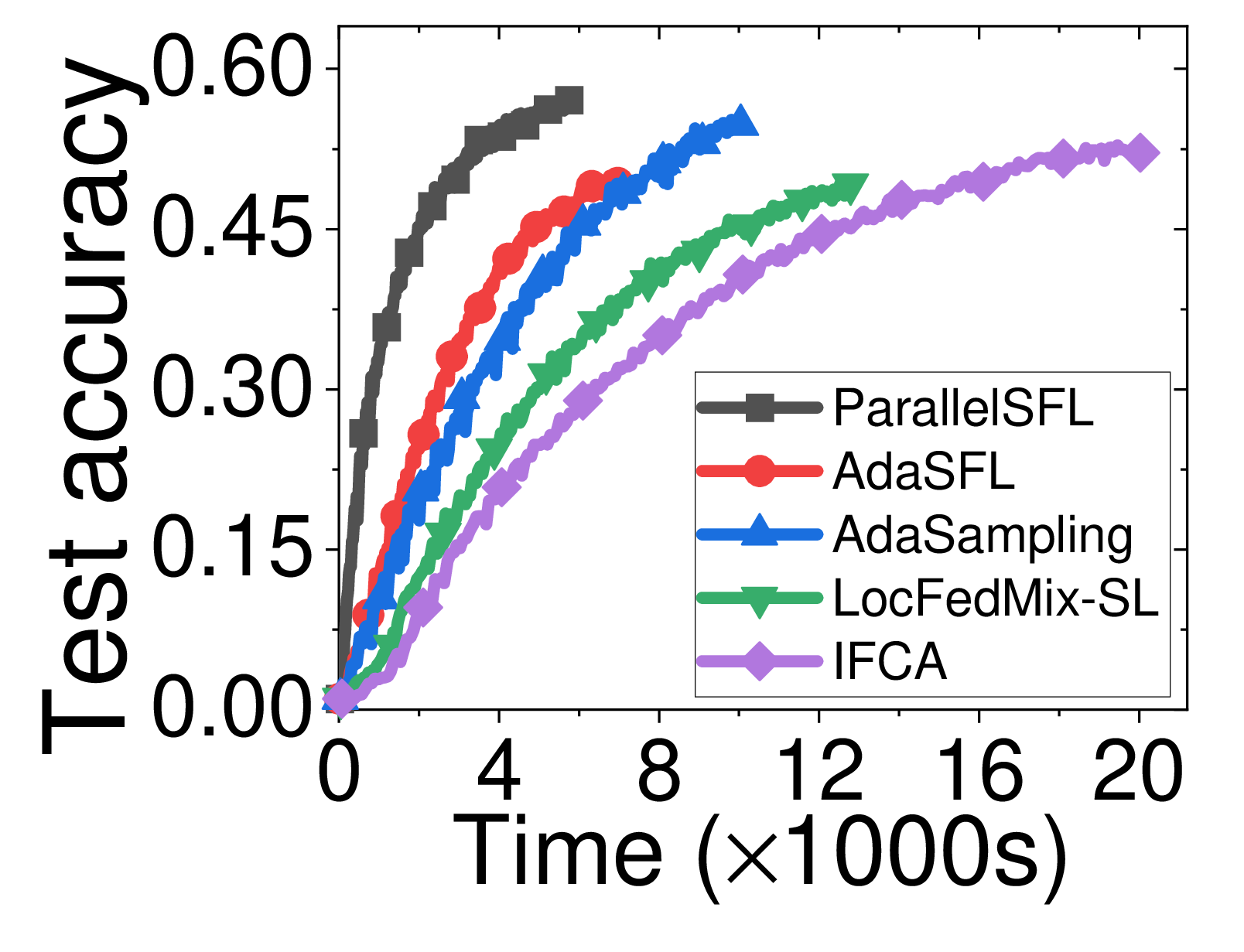}
    \label{fig:acc_noniid_image}
}\quad
\subfigure[QNLI]
{
    \includegraphics[width=0.225\linewidth,height=3.3cm]{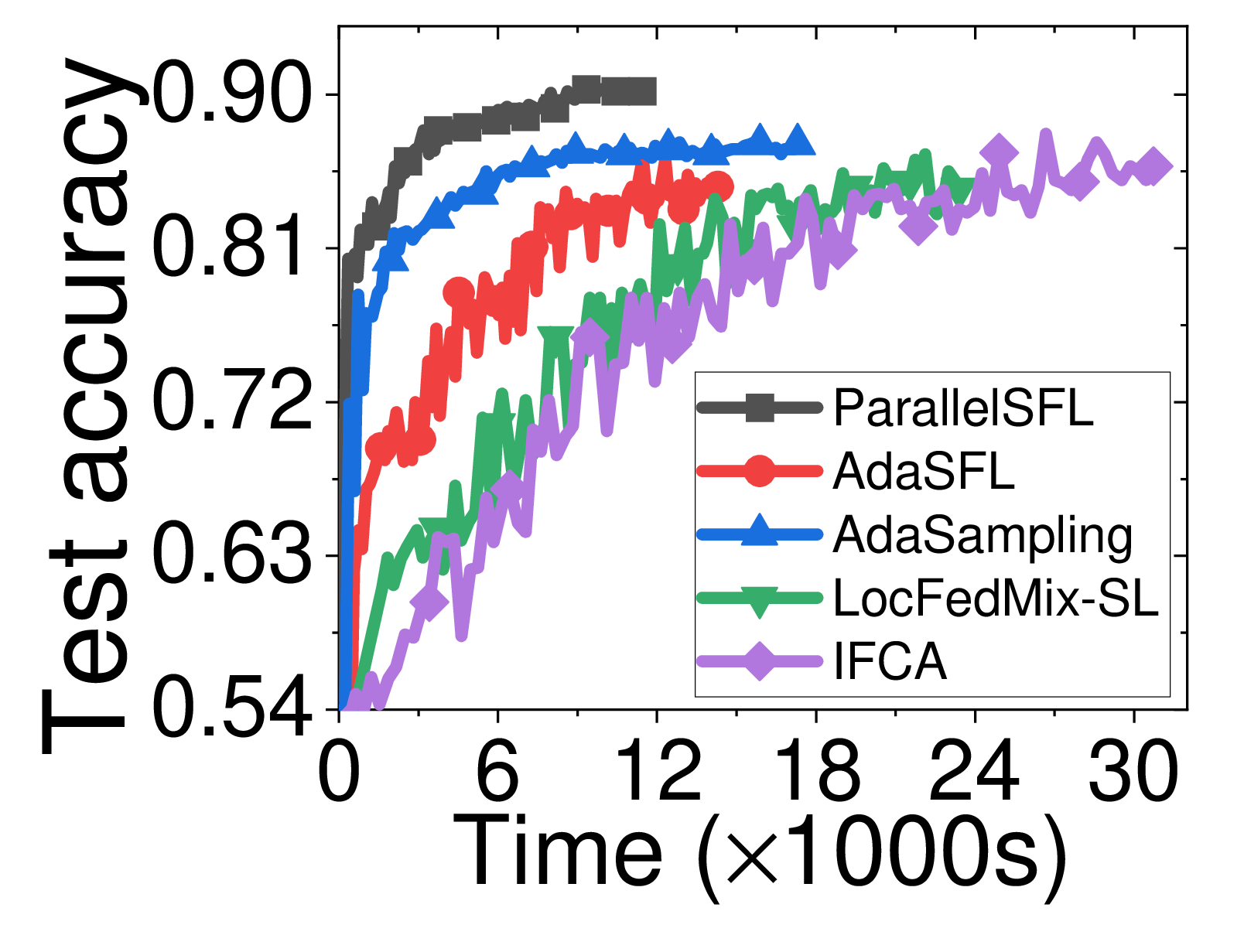}
    \label{fig:acc_noniid_Bert}
}
\caption{Test accuracy of five approaches on the four non-IID datasets.}
\label{fig:acc_noniid}
\end{figure*}

\textbf{Baselines.}
We measure the effectiveness of \method through a comparison with four approaches.


\bluenote{
1) \textbf{\textit{IFCA}} \cite{sattler2020clustered} is a famous clustered FL approach that alternates between estimating the cluster identities and minimizing the loss functions to deal with statistical heterogeneity.}

2) \textbf{\textit{LocFedMix-SL}} \cite{oh2022locfedmix} is a typical SFL approach, which proposes to reduce the aggregation frequency of bottom submodels to save the traffic consumption, but can not fully utilize the capacities of heterogeneous workers.

3) \textbf{\textit{AdaSampling}} \cite{luo2022tackling} is an advanced FL approach that focuses on designing an adaptive
worker sampling algorithm to tackle both system as well as statistical heterogeneity and minimize the wall-clock convergence time.

4) \textbf{\textit{AdaSFL}} \cite{liao2023accelerating} is a state-of-the-art SFL approach, which assigns adaptive local updating frequency and diverse batch sizes for heterogeneous workers to enhance the training efficiency without addressing statistical heterogeneity.

\textbf{Metrics.}
We adopt the following metrics to evaluate the performance of \method and the baselines.

1) \textbf{\textit{Test Accuracy}}  reflects the accuracy of the models trained by different approaches on the test datasets, and is measured by the proportion of the data correctly predicted by the models to all the test data.
We evaluate the test accuracy of the global model in each round, and record the final accuracy for all approaches.

2) \textbf{\textit{Training Time}} is denoted as the total wall clock time taken for training a model to achieve a target accuracy.
For a fair comparison, we set the target accuracy as the achievable accuracy by all approaches.
In addition, we also record the average waiting time of all workers to reflect the training efficiency of different approaches.

3) \textbf{\textit{Network Traffic}} is calculated by summing the traffic for transmitting models or features between the PS and workers or between the bottom workers and top workers when achieving a target accuracy.

\textbf{Experimental Parameters.}
\bluenote{
By default, each set of experiments will run 100 aggregation rounds for RoBERTa on QNLI, and 250 aggregation rounds for CNN on Speech, AlexNet on CIFAR-10, and VGG16 on IMAGE-100.
The learning rates and decay rates for CNN, AlexNet, and VGG16 are identical, and are initialized as 0.1 and 0.993 \cite{liao2023adaptive, rothchild2020fetchsgd}, respectively, while for RoBERTa, the learning rate is initialized as 0.001 without decay rate.
Besides, the batch size is set as 16 for RoBERTa and 64 for the remaining three models.
For the SFL approaches, we separately split the CNN, AlexNet, VGG16, and RoBERTa at the 4th, 5th, 13th, and 3rd layer \cite{liao2023accelerating}.}


\subsection{Overall Performance}
\vspace{0.15cm}

Firstly, we conduct a set of experiments on the four IID datasets to evaluate the performance of \method and the baselines.
The training processes of five approaches are presented in Fig. \ref{fig:acc_iid}.
By the results, all the approaches achieve similar test accuracy eventually on the four datasets.
However, \method achieves the fastest convergence, followed by AdaSFL, which is much faster than the other approaches on all the four datasets.
\bluenote{
For instance, as illustrated in Fig. \ref{fig:acc_iid_speech}, \method takes 2,893s to achieve 87\% accuracy for CNN on Speech, while AdaSFL, AdaSampling, LocFedMix-SL, and IFCA consume 3,267s, 4,877s, 6,878s and 9,253s, respectively.
Similarly, by Fig. \ref{fig:acc_iid_CIFAR10}, for AlexNet on CIFAR-10, \method can separately speed up training by about 1.36$\times$, 1.65$\times$, 2.18$\times$ and 2.87$\times$, compared to AdaSFL, AdaSampling, LocFedMix-SL, and IFCA, respectively.
AdaSFL with adaptive and diverse batch sizes for heterogeneous workers speeds up the convergence rate, while \method with adaptive worker clustering and local updating frequency optimization overcomes the system heterogeneity and achieves the fastest convergence.
Specifically, for VGG16 on IMAGE-100, as shown in Fig. \ref{fig:acc_iid_image}, \method reduces the total training time by about 17\%, 43\%, 55\%, and 68\%, compared to the baselines (\ie, AdaSFL, AdaSampling, LocFedMix-SL, and IFCA).
Moreover, by Fig. \ref{fig:acc_iid_Bert}, \method takes 10,838s to achieve 92\% accuracy for RoBERTa on QNLI, while AdaSFL, AdaSampling, LocFedMix-SL, and IFCA consume 13,635s, 16,812s, 22,565s, 29,190s, respectively.
}

\begin{figure*}[!t]
\centering
\subfigure[Speech]
{
    \includegraphics[width=0.225\linewidth,height=3.3cm]{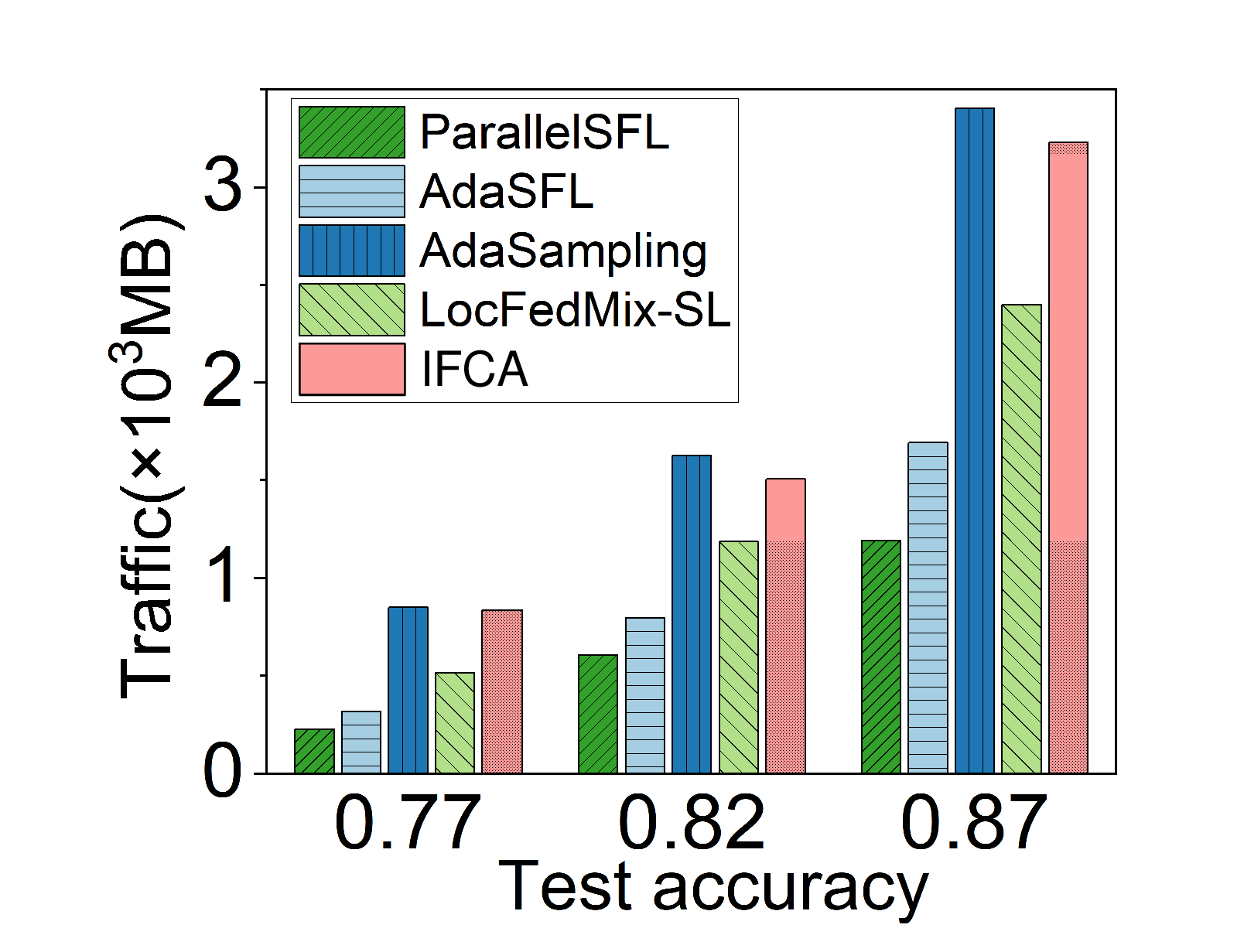}
    \label{fig:bandwidth_speech}
}\quad 
\subfigure[CIFAR-10]
{
    \includegraphics[width=0.225\linewidth,height=3.3cm]{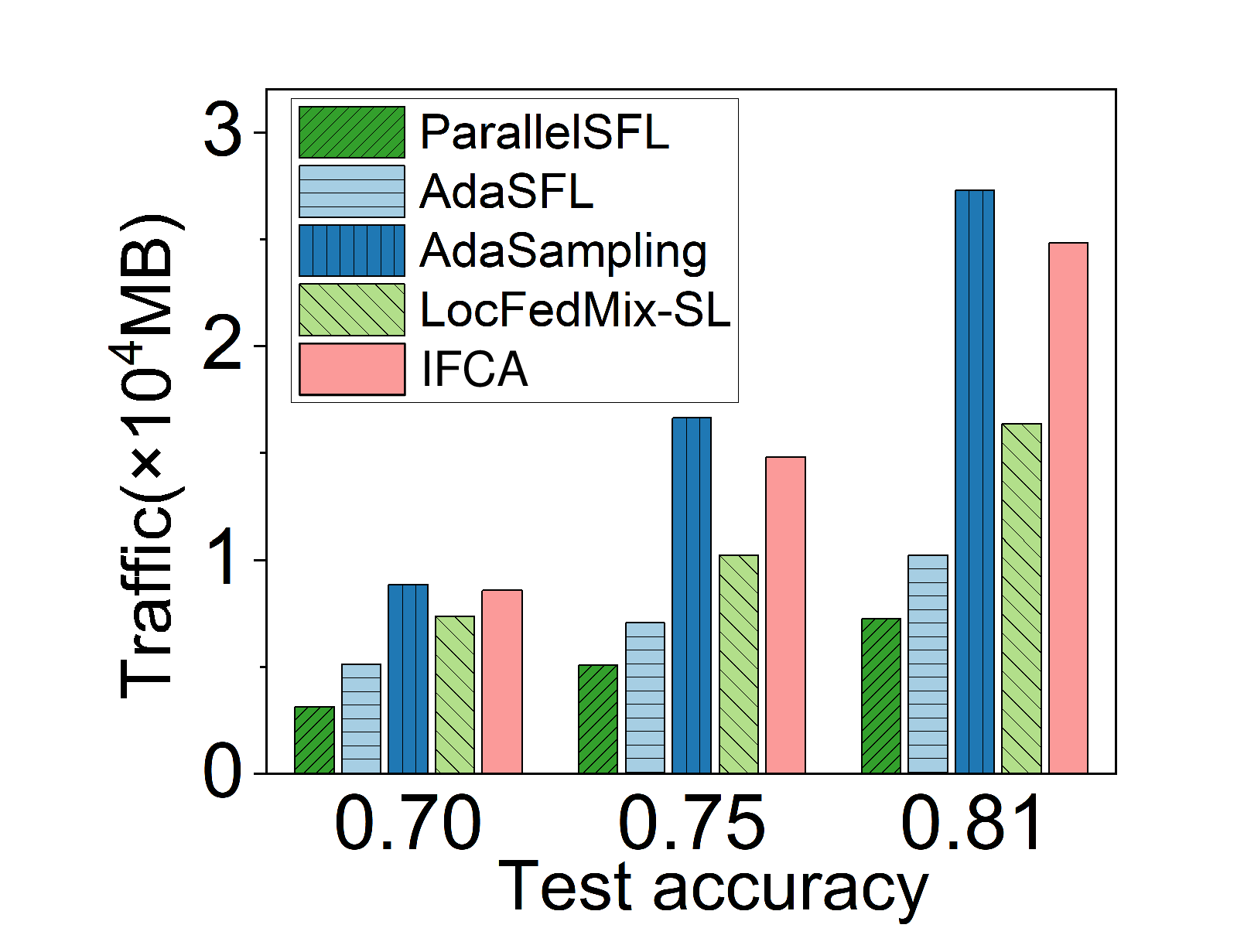}
    \label{fig:bandwidth_CIFAR10}
}\quad 
\subfigure[IMAGE-100]
{
    \includegraphics[width=0.225\linewidth,height=3.3cm]{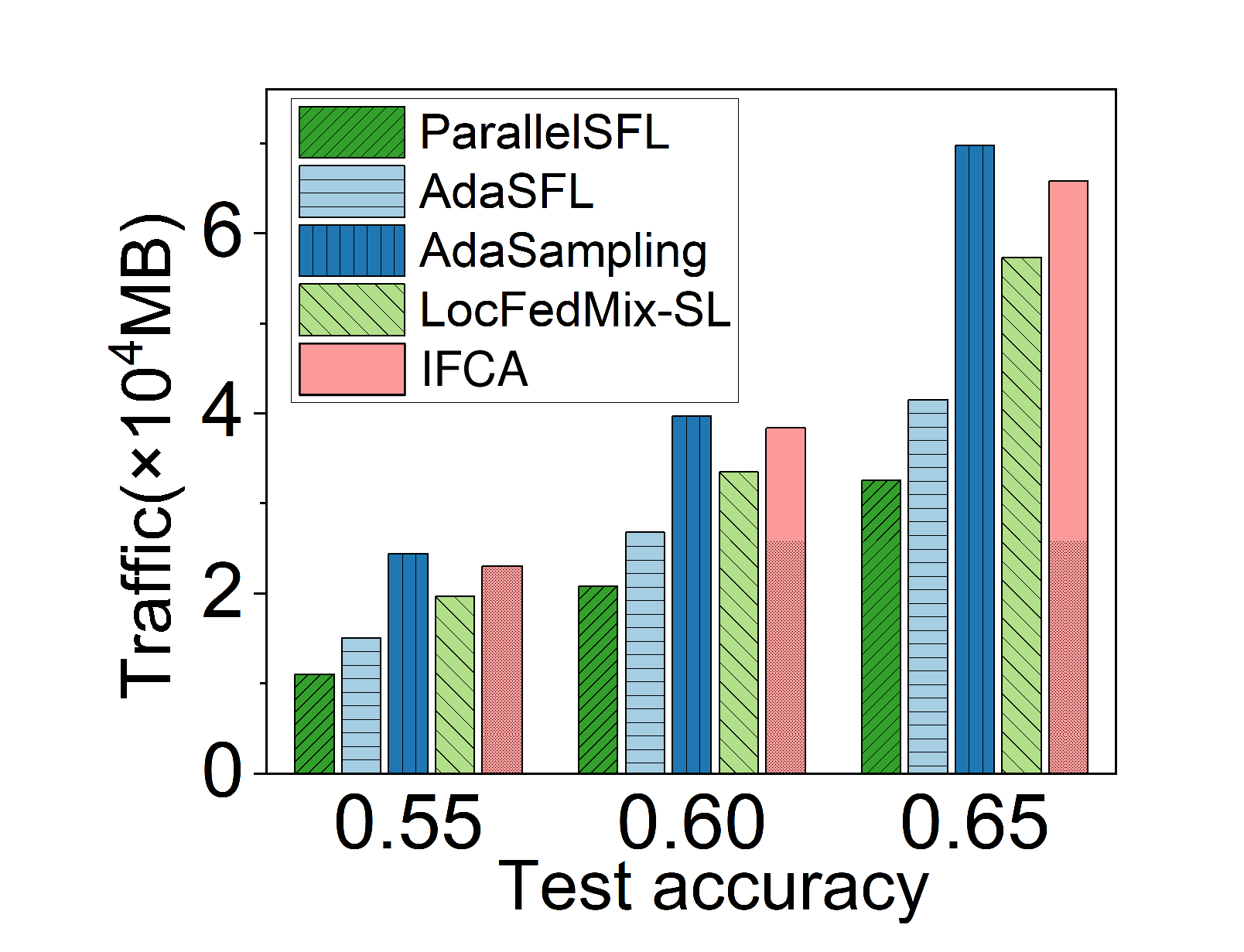}
    \label{fig:bandwidth_image}
}\quad
\subfigure[QNLI]
{
    \includegraphics[width=0.225\linewidth,height=3.3cm]{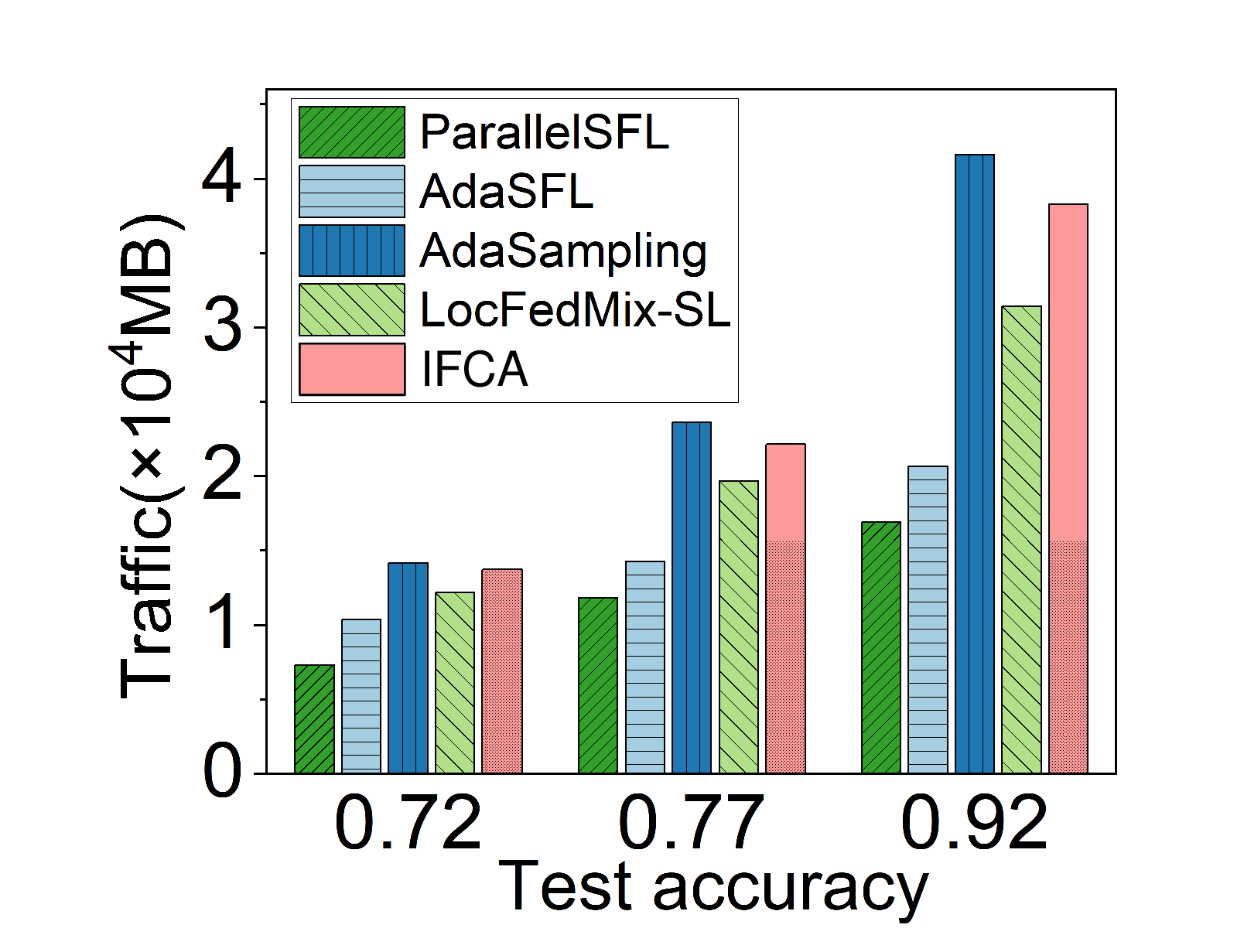}
    \label{fig:bandwidth_Bert}
} 
\caption{Network traffic consumption of five approaches when achieving different target accuracies.}
\label{fig:bandwidth}
\end{figure*}

\begin{figure*}[!t]
\centering
\subfigure[Speech]
{
    \includegraphics[width=0.225\linewidth,height=3.3cm]{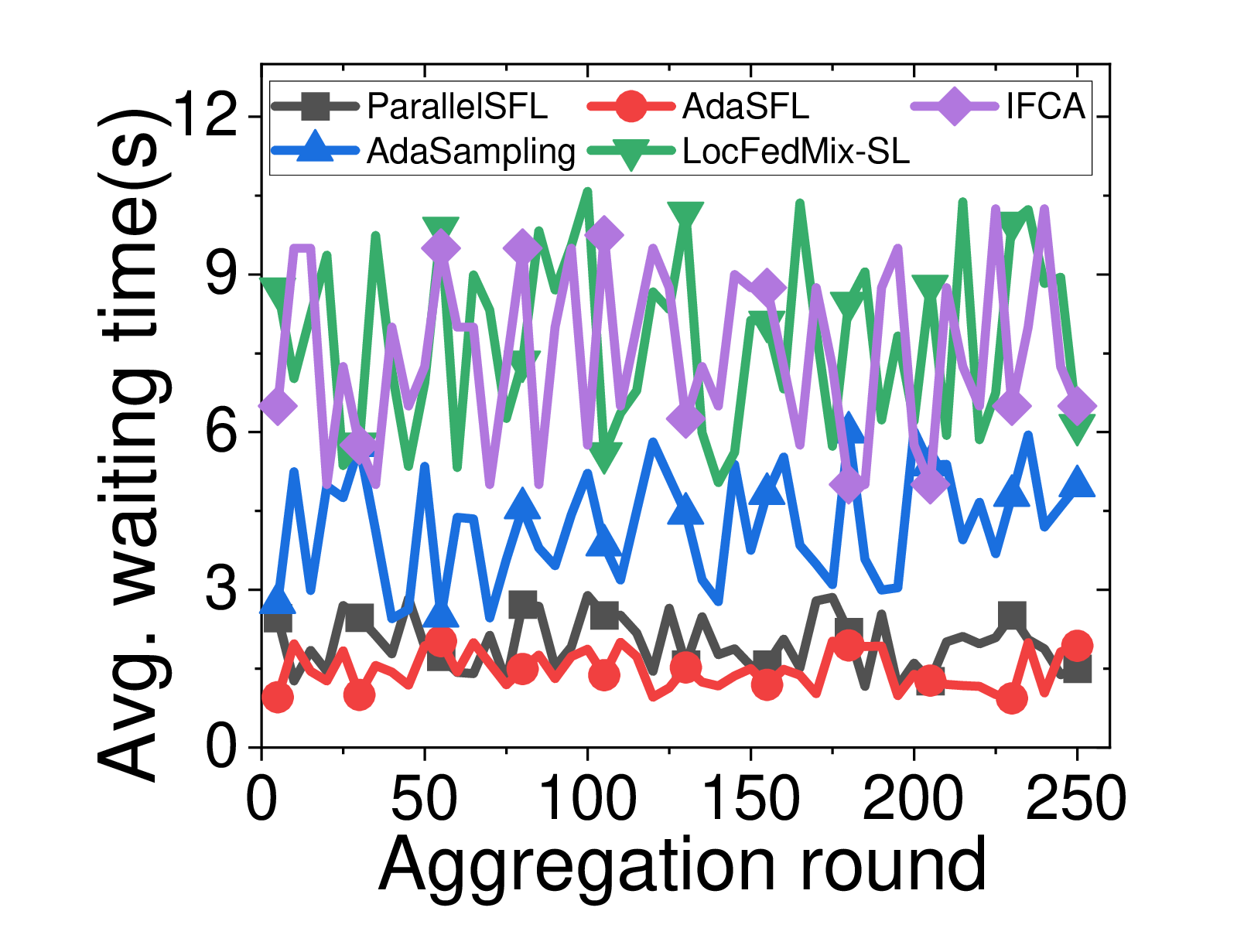}
    \label{fig:waiting_time_speech}
}\quad 
\subfigure[CIFAR-10]
{
    \includegraphics[width=0.225\linewidth,height=3.3cm]{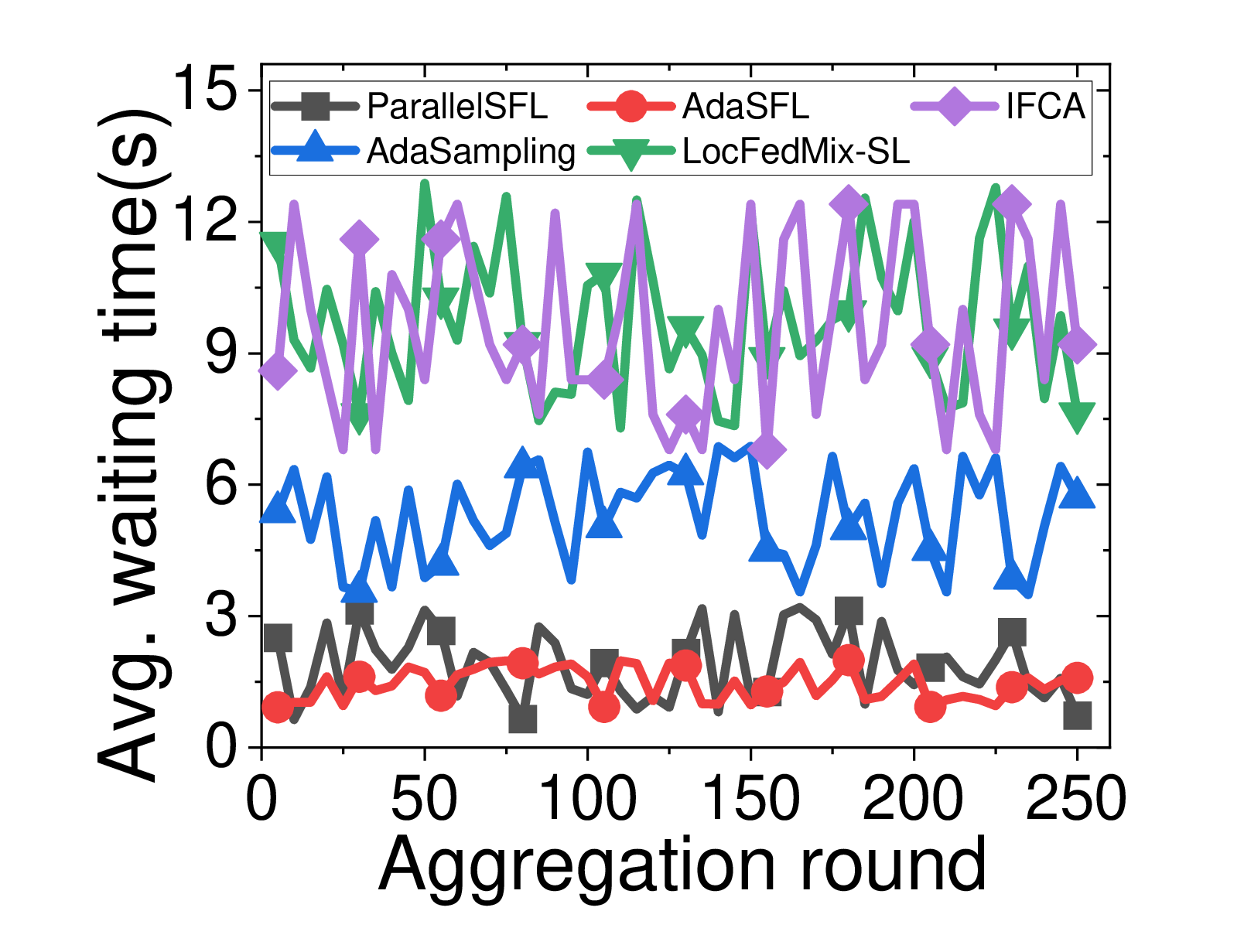}
    \label{fig:waiting_time_CIFAR10}
}\quad 
\subfigure[IMAGE-100]
{
    \includegraphics[width=0.225\linewidth,height=3.3cm]{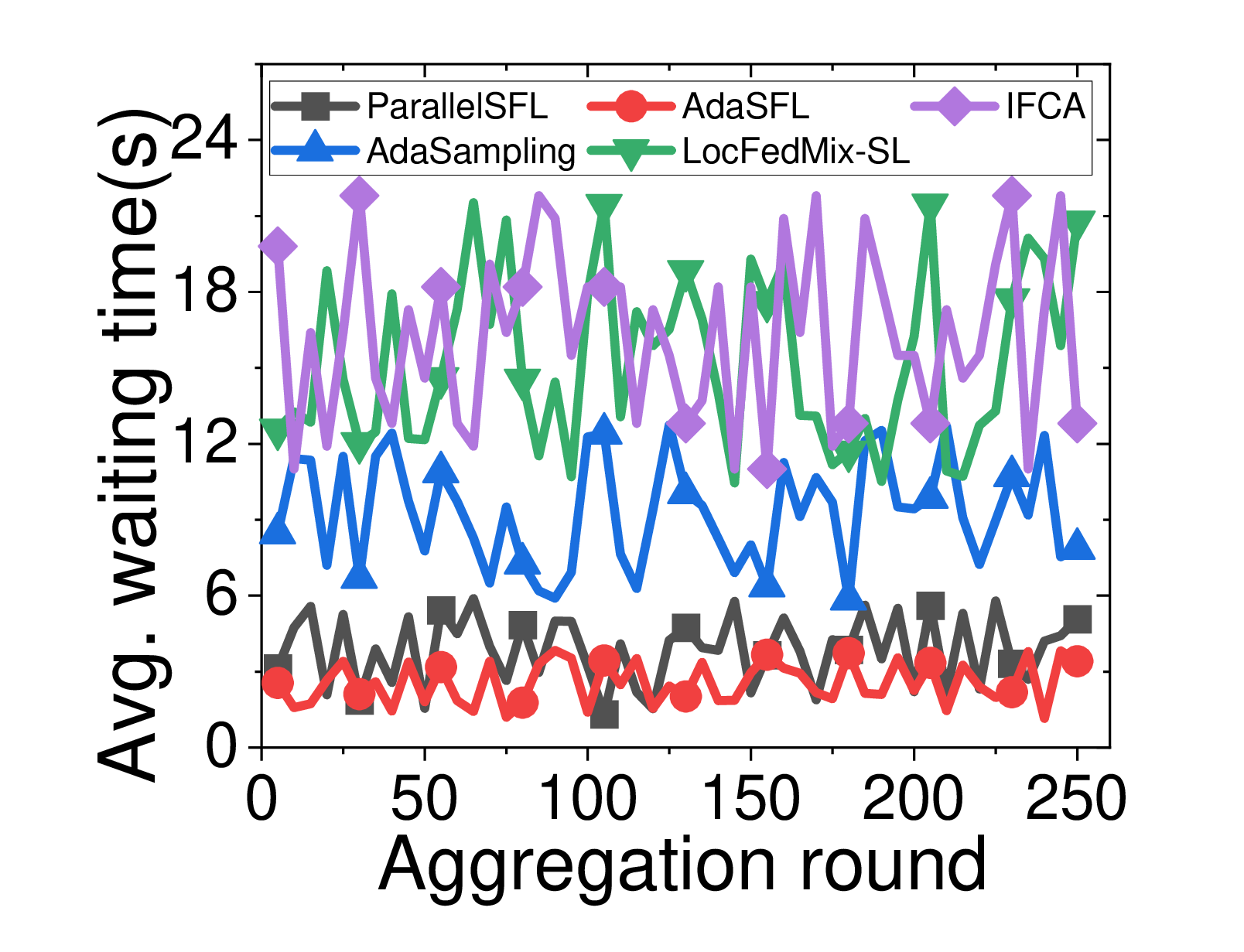}
    \label{fig:waiting_time_image}
}\quad
\subfigure[QNLI]
{
    \includegraphics[width=0.225\linewidth,height=3.3cm]{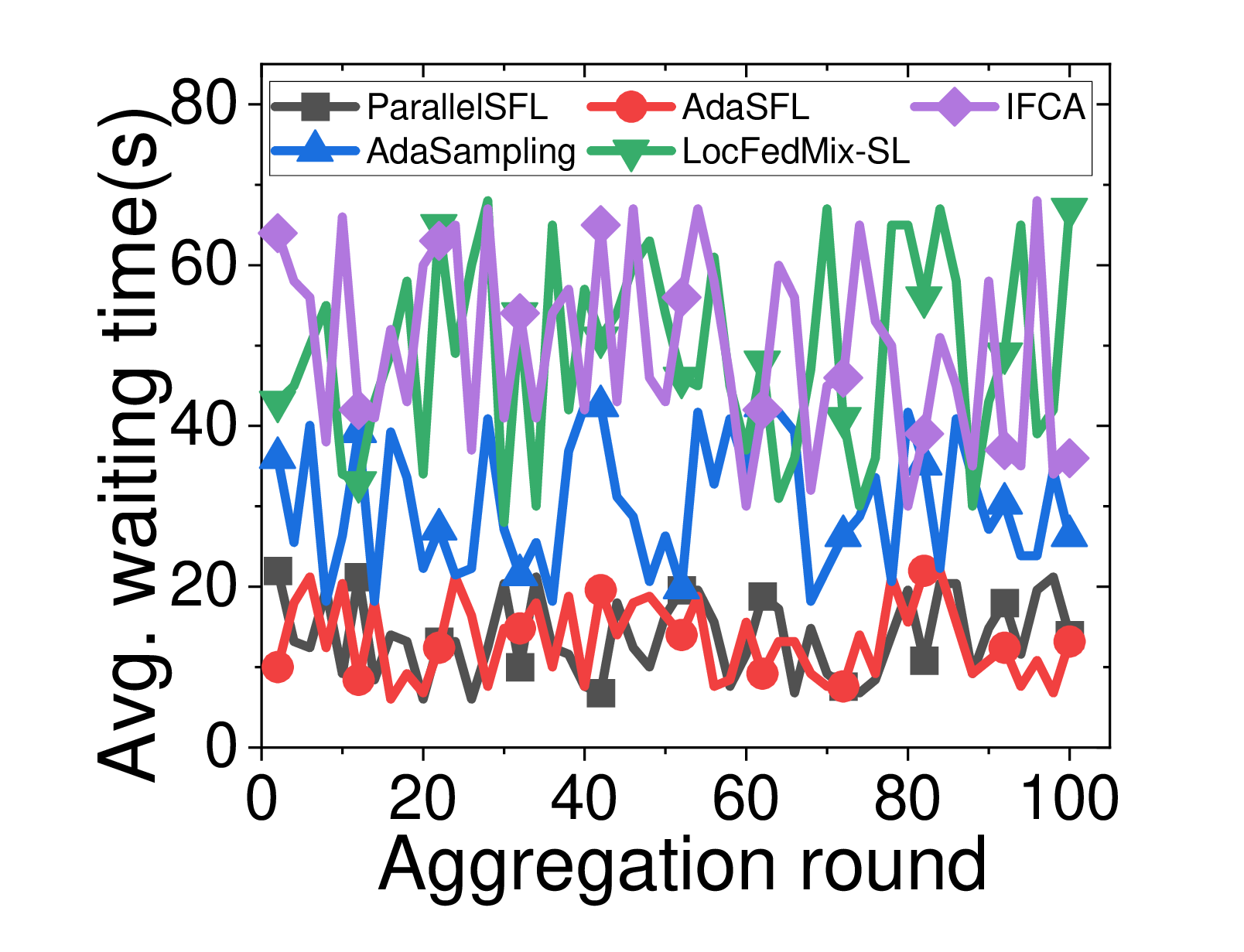}
    \label{fig:waiting_time_Bert}
}
\caption{Average waiting time of five approaches on the four datasets.}
\label{fig:waiting_time}
\end{figure*}

Secondly, we also conduct a set of experiments of these approaches on the four datasets with non-IID level $p$=10, and the results are presented in Fig. \ref{fig:acc_noniid}. 
We observe that all the approaches maintain a similar convergence rate as that in the IID scenario, but suffer from varying degrees of accuracy degradation.
However, \method with adaptive worker clustering and model splitting achieves the highest accuracy among these approaches.
\bluenote{
For instance, by Fig. \ref{fig:acc_noniid_speech}, \method achieves 82.1\% accuracy in 2,972s for CNN on Speech, while AdaSFL, AdaSampling, LocFedMix-SL, and IFCA takes 3,567s, 5,747s, 7,737s, and 9,885s to reach the accuracy of 68.64\%, 80.07\%, 68.23\%, and 71.31\%, respectively.
Similarly, as shown in Fig. \ref{fig:acc_noniid_CIFAR10}, for AlexNet on CIFAR-10 with the same training time of 4,300s, \method improves the test accuracy by about 24.12\%, 19.08\%, 36.45\% and 37.28\%, compared to AdaSFL, AdaSampling, LocFedMix-SL and IFCA, respectively.
IFCA with alternate worker clustering improves the accuracy to a certain extent, while AdaSampling with adaptive worker sampling speeds up the convergence rate and improves test accuracy.
However, \method with adaptive worker clustering and model splitting tackles both system and statistical heterogeneity better than IFCA and AdaSampling.
Specifically, Fig. \ref{fig:acc_noniid_image} illustrates that \method separately improves the final accuracy by about 15.7\%, 5.74\%, 16.06\% and 10.63\% for VGG16 on IMAGE-100, compared to AdaSFL, AdaSampling, LocFedMix-SL, and IFCA, respectively.
Moreover, as shown in Fig. \ref{fig:acc_noniid_Bert}, \method achieves 90.11\% accuracy in 11,445s for RoBERTa on QNLI, while AdaSFL, AdaSampling, LocFedMix-SL, and IFCA take 14,293s, 17,306s, 23,403s, and 30,724s to reach the accuracy of 84.51\%, 87.33\%, 84.62\%, and 85.76\%, respectively.
These results demonstrate that \method is effective in simultaneously tackling the system and statistical heterogeneity.
}

Thirdly, to illustrate the advantage of \method in saving communication resource, we illustrate the network traffic consumption of these approaches when achieving different target accuracies in Fig. \ref{fig:bandwidth}.
By the results, the network traffic consumption of all approaches increases with the target accuracy for all the four datasets.
Furthermore, \method always consumes the least network traffic among all approaches.
In addition, model splitting (\ie, \method, AdaSFL and LocFedMix-SL) helps to save much more network traffic compared to typical FL approaches (\ie, IFCA and AdaSampling).
AdaSFL with adaptive local updating frequency reduces the network traffic consumption while \method with adaptive worker clustering further reduces the network traffic consumption.
\bluenote{
Specifically, As shown in Fig. \ref{fig:bandwidth_speech}, when achieving 87\% accuracy, \method, AdaSFL and LocFedMix-SL consume 1,192MB, 1,692MB and 2,398MB, respectively, while AdaSampling and IFCA consume 3,403MB and 3,228MB for CNN on Speech.
By Fig. \ref{fig:bandwidth_CIFAR10}, for training AlexNet on CIFAR-10 to achieve 81\% accuracy, \method reduces the network traffic consumption by about 3,012MB, 20,018MB, 9,116MB, and 17,564MB, compared to AdaSFL, AdaSampling, LocFedMix-SL, and IFCA, respectively.
Besides, as illustrated in Fig. \ref{fig:bandwidth_image}, \method saves network traffic consumption by about 21\%, 53\%, 43\%, and 49\% when achieving 65\% accuracy for VGG16 on IMAGE-100, compared to the baselines (\ie, AdaSFL, AdaSampling, LocFedMix-SL, and IFCA).
Moreover, by Fig. \ref{fig:bandwidth_Bert}, for training RoBERTa on QNLI to achieve 92\% accuracy, \method reduces the network traffic consumption by about 3,783MB, 24,747MB, 14,537MB, and 21,422MB, compared to AdaSFL, AdaSampling, LocFedMix-SL, and IFCA, respectively.
These results demonstrate that \method is effective in saving network traffic consumption.
}

\begin{figure*}[!t]
\centering
\subfigure[Speech]
{
    \includegraphics[width=0.225\linewidth,height=3.3cm]{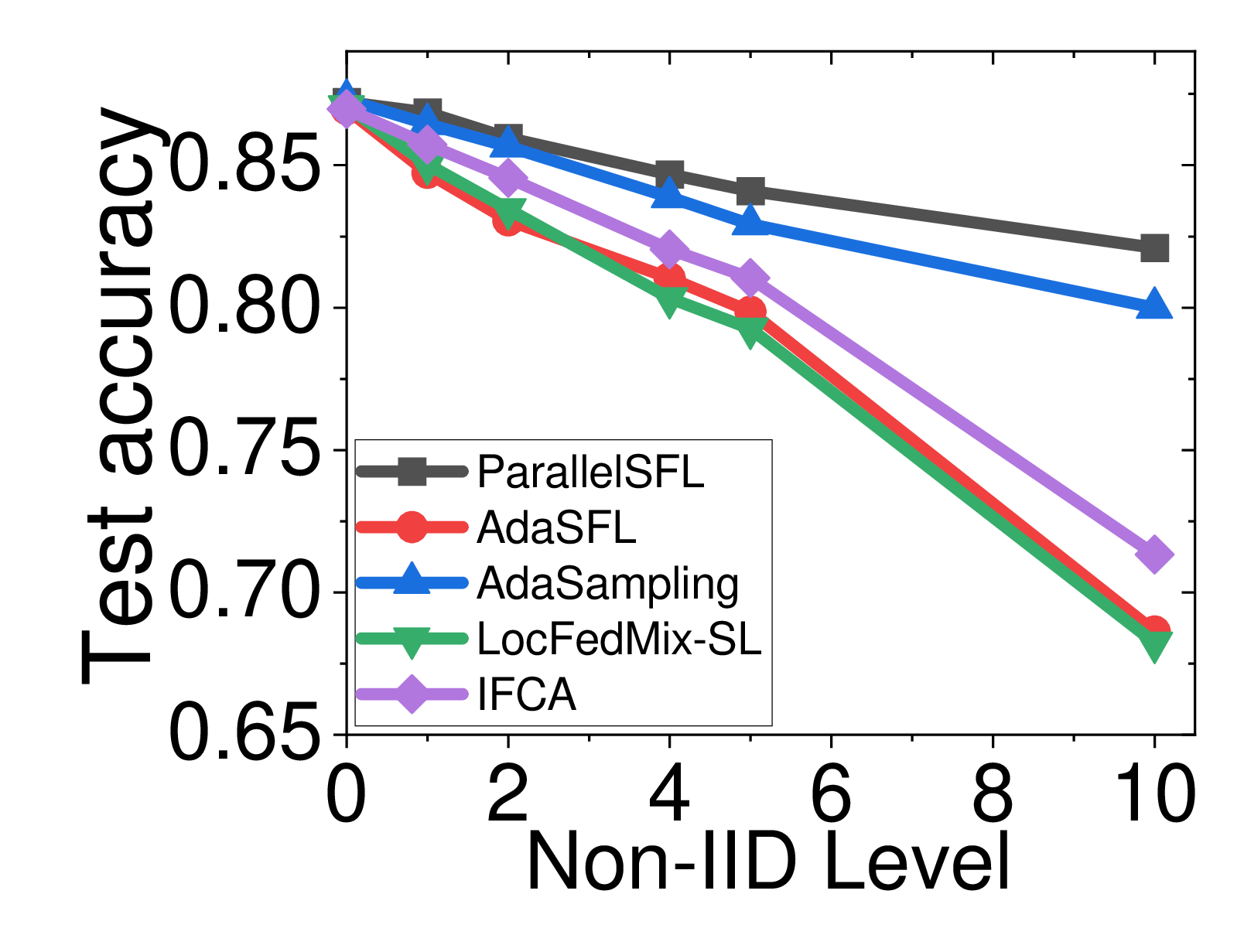}
    \label{fig:noniid_level_speech}
}\quad 
\subfigure[CIFAR-10]
{
    \includegraphics[width=0.225\linewidth,height=3.3cm]{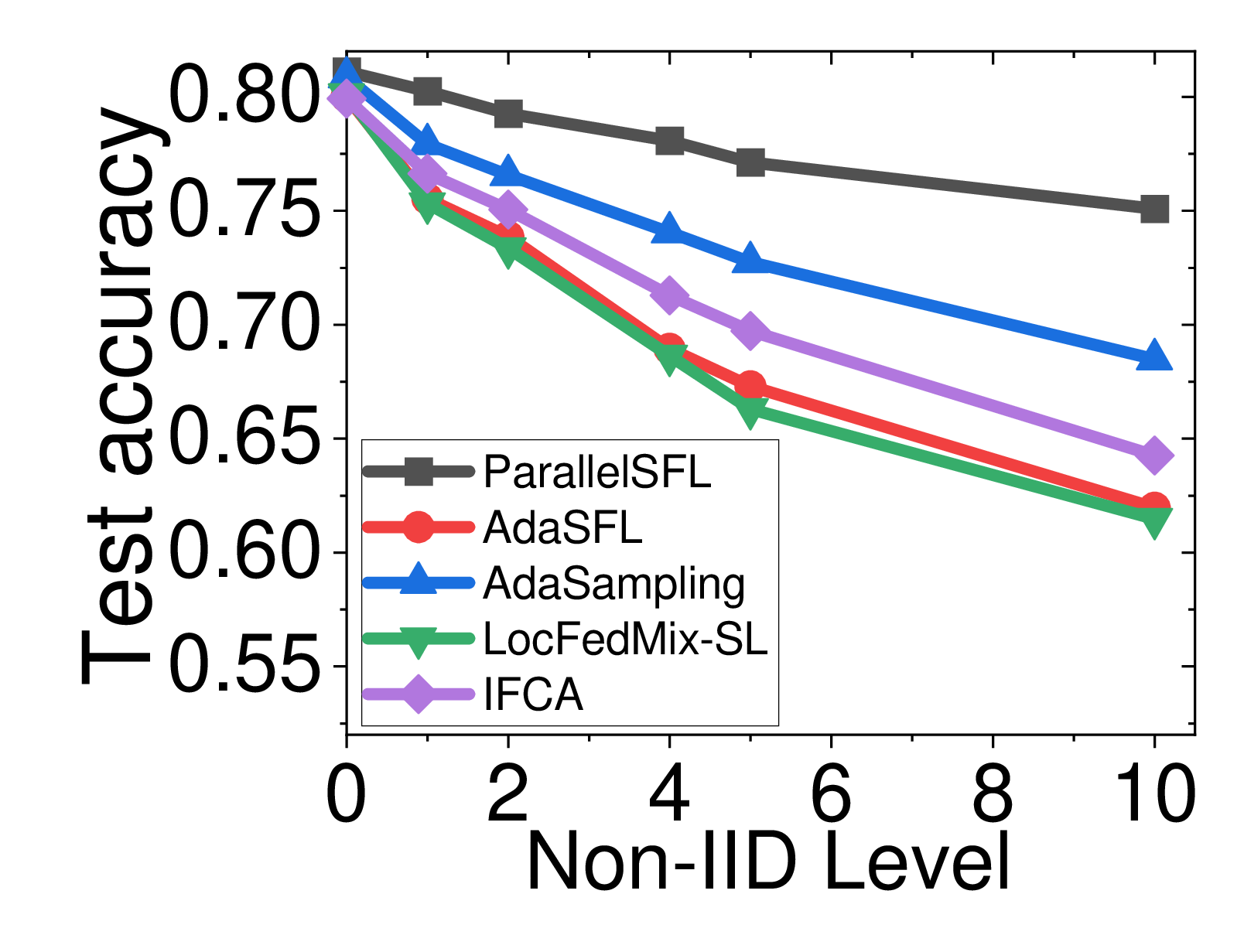}
    \label{fig:noniid_level_CIFAR10}
}\quad 
\subfigure[IMAGE-100]
{
    \includegraphics[width=0.225\linewidth,height=3.3cm]{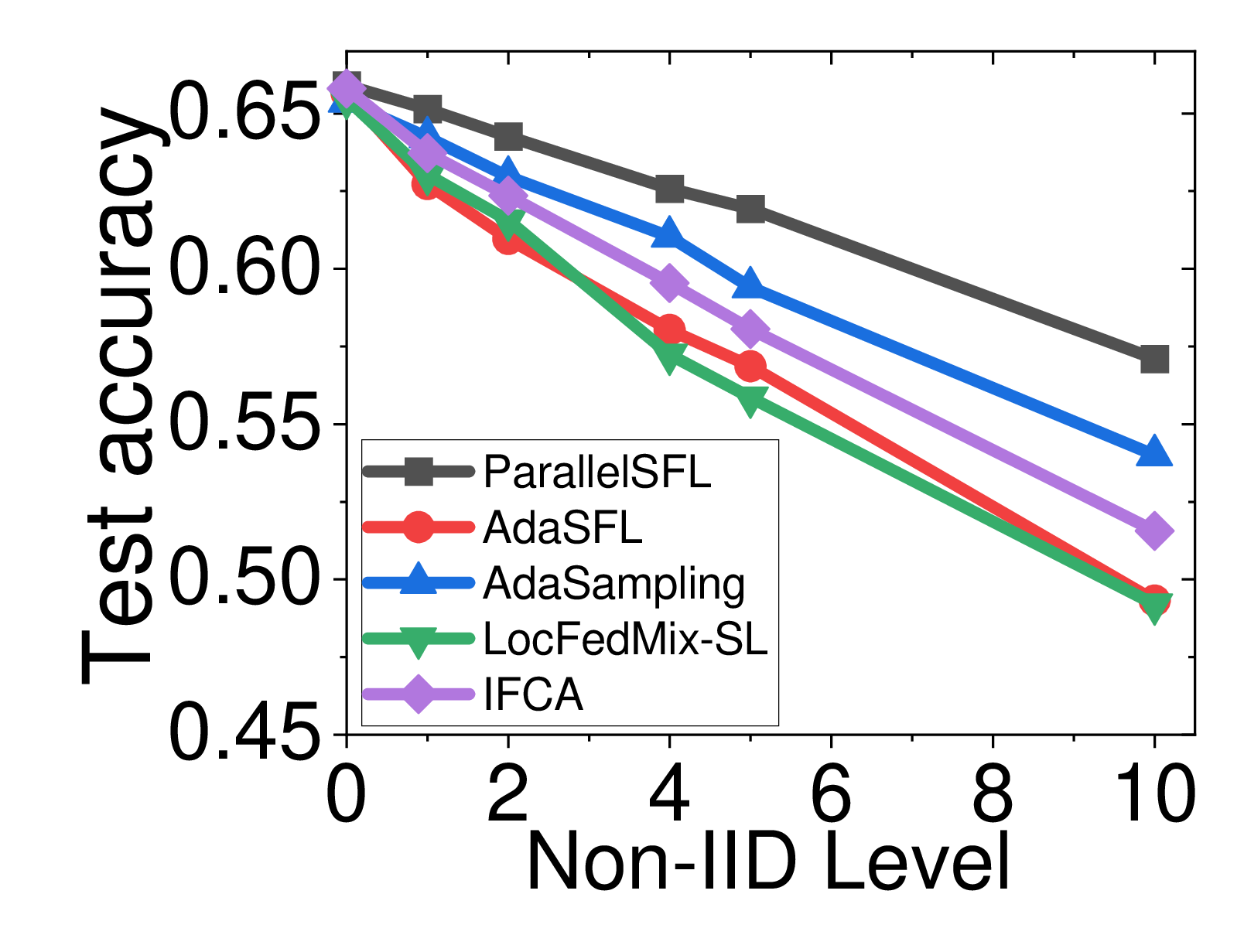}
    \label{fig:noniid_level_image}
}\quad
\subfigure[QNLI]
{
    \includegraphics[width=0.225\linewidth,height=3.3cm]{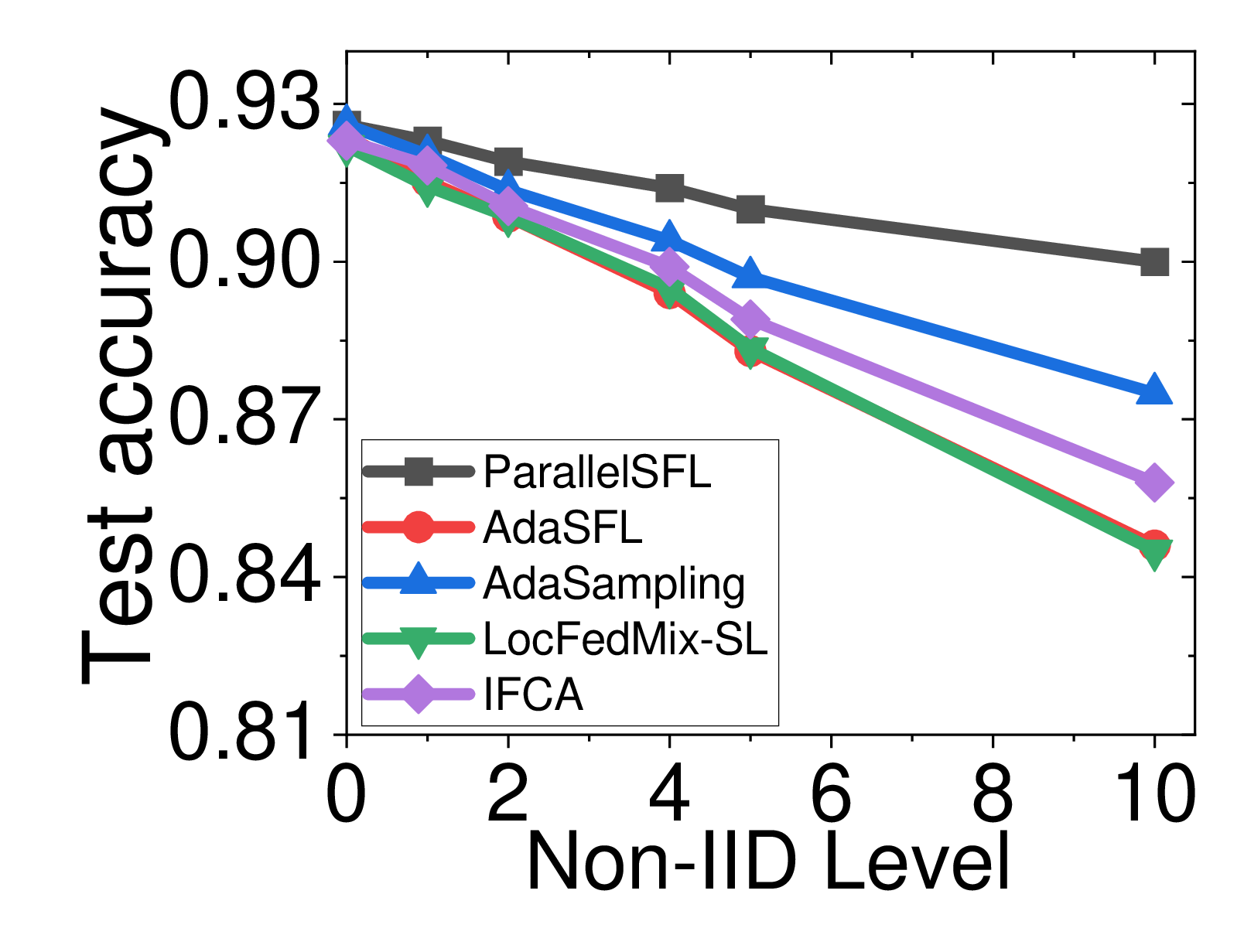}
    \label{fig:noniid_level_Bert}
}
\caption{Test accuracy varies with different non-IID levels.}
\label{fig:noniid_level}
\end{figure*}

\begin{figure}[!t]
\centering
\subfigure[IID]
{
    \includegraphics[width=0.45\linewidth,height=3.3cm]{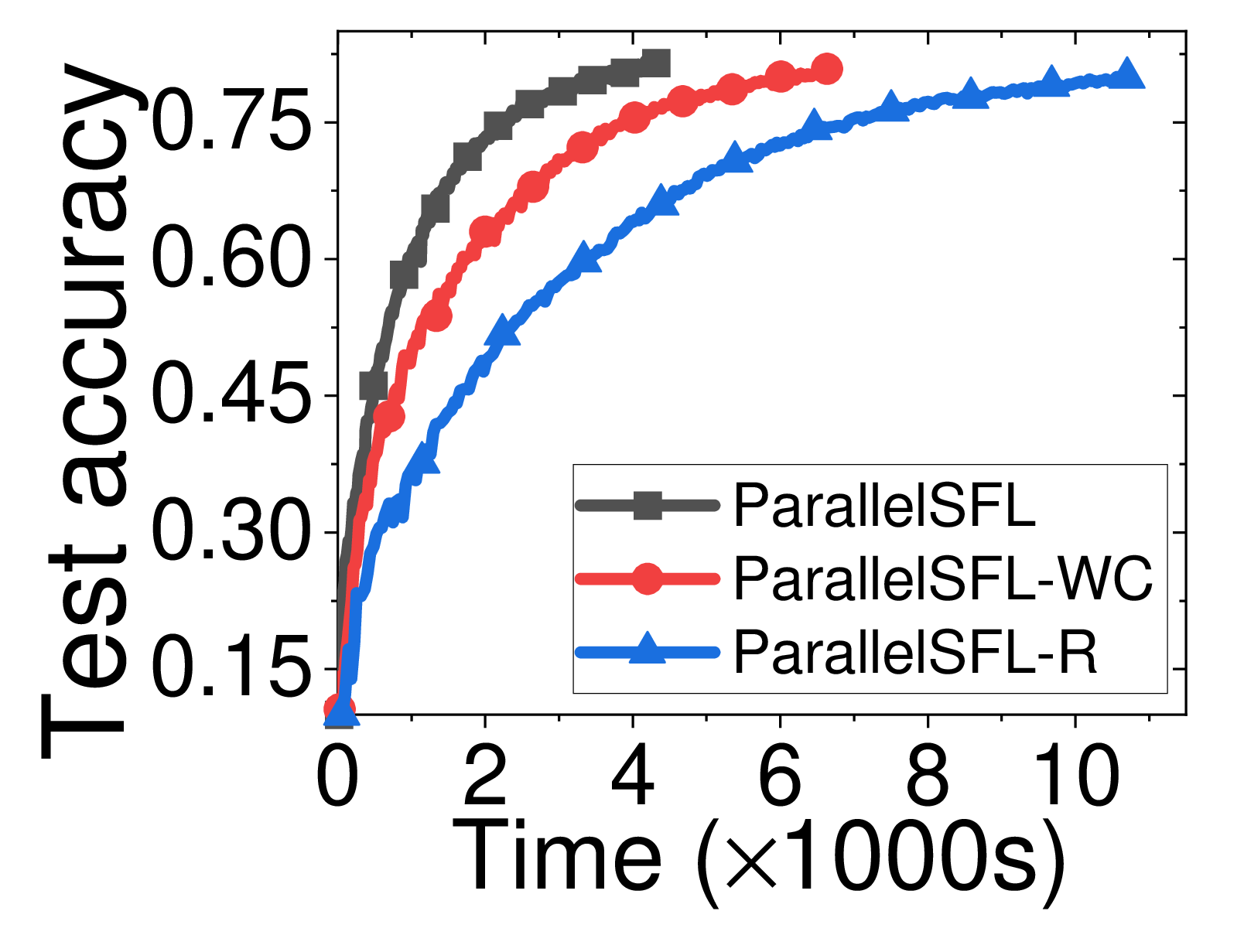}
    \label{fig:key_IID}
}\quad
\subfigure[Non-IID]
{
    \includegraphics[width=0.45\linewidth,height=3.3cm]{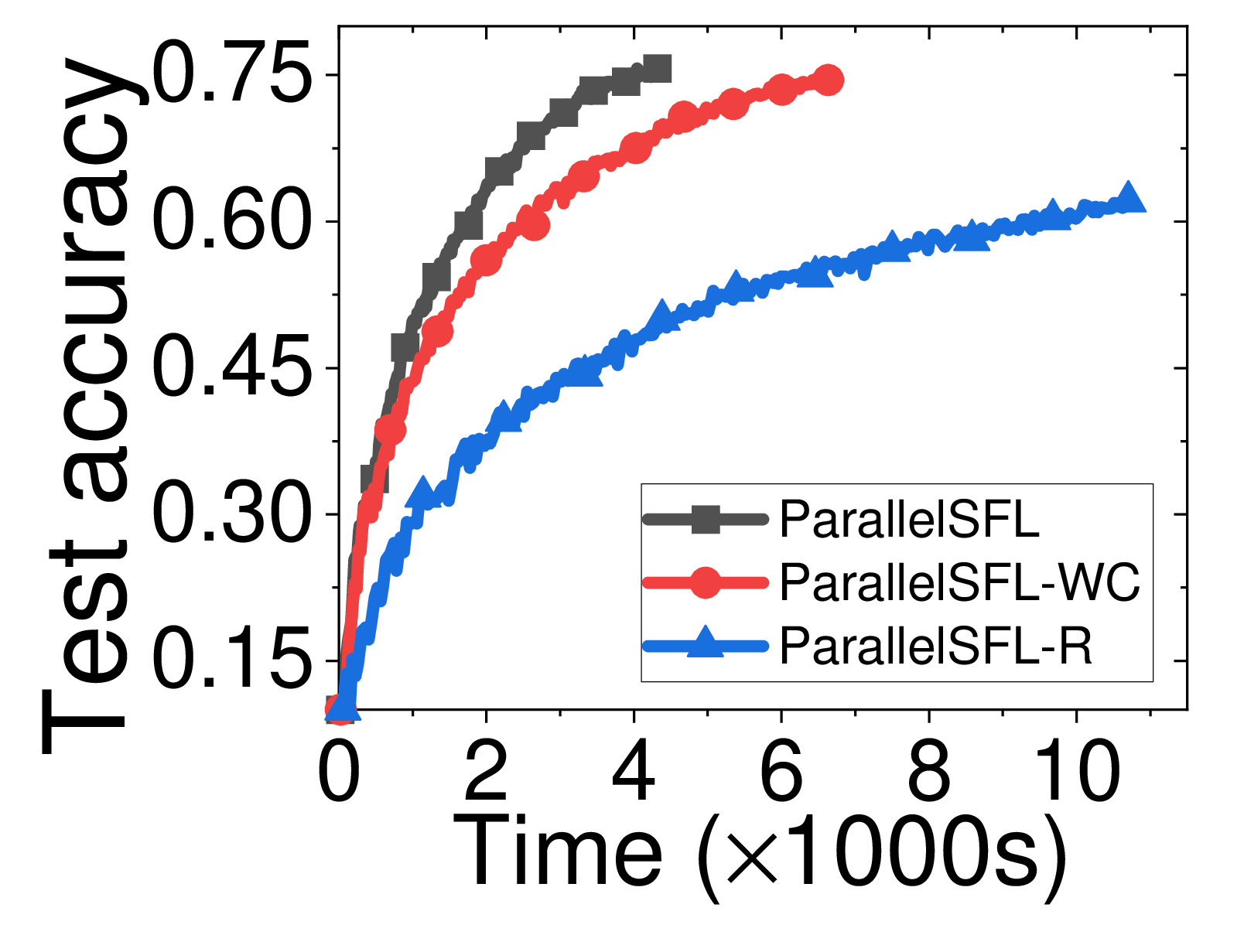}
    \label{fig:key_nonIID}
}
\caption{Effects of worker clustering and local updating frequency optimization.}
\label{fig:key_strategies}
\end{figure}

Finally, to further demonstrate the robustness of \method towards system heterogeneity, we illustrate the average waiting time of the five approaches on both datasets in Fig. \ref{fig:waiting_time}.
AdaSFL with adaptive and diverse batch sizes for heterogeneous workers achieves the least waiting time, but the waiting time of \method is close to AdaSFL and is much less than that of other approaches.
\bluenote{
For instance, as shown in Fig. \ref{fig:waiting_time_speech}, the average waiting time of \method is 1.5s for CNN on Speech while AdaSFL, AdaSampling, LocFedMix-SL, and IFCA incur that of 1.3s, 4.4s, 7.3s and 7.1s, respectively.
Similarly, by Fig. \ref{fig:waiting_time_CIFAR10}, compared to AdaSampling, LocFedMix-SL and IFCA, \method reduces the average waiting time to train AlexNet on CIFAR-10 by about 53\%, 67\% and 62\%, respectively.
Considering the workers with varying capacities, LocFedMix-SL and IFCA do not consider system heterogeneity, thus leading to non-negligible waiting time.
AdaSampling with adaptive worker sampling reduces the average waiting time to a certain extent.
Specifically, by Fig. \ref{fig:waiting_time_image}, \method reduces the average waiting time for VGG16 on IMAGE-100 by about 63\%, 80\%, and 79\%, compared to AdaSampling, LocFedMix-SL, and IFCA, respectively.
Moreover, by Fig. \ref{fig:waiting_time_Bert}, the average waiting time of \method is 10.1s for RoBERTa on QNLI while AdaSampling, AdaSFL, LocFedMix-SL, and IFCA incur an average waiting time of 9.6s, 34.5s, 53.7s, and 51.3s, respectively.
In general, these results illustrate that \method overcomes the challenge of system heterogeneity well, compared to existing methods.
}

\begin{figure}[!t]
\centering
\subfigure[Completion Time]
{
    \includegraphics[width=0.45\linewidth,height=3.3cm]{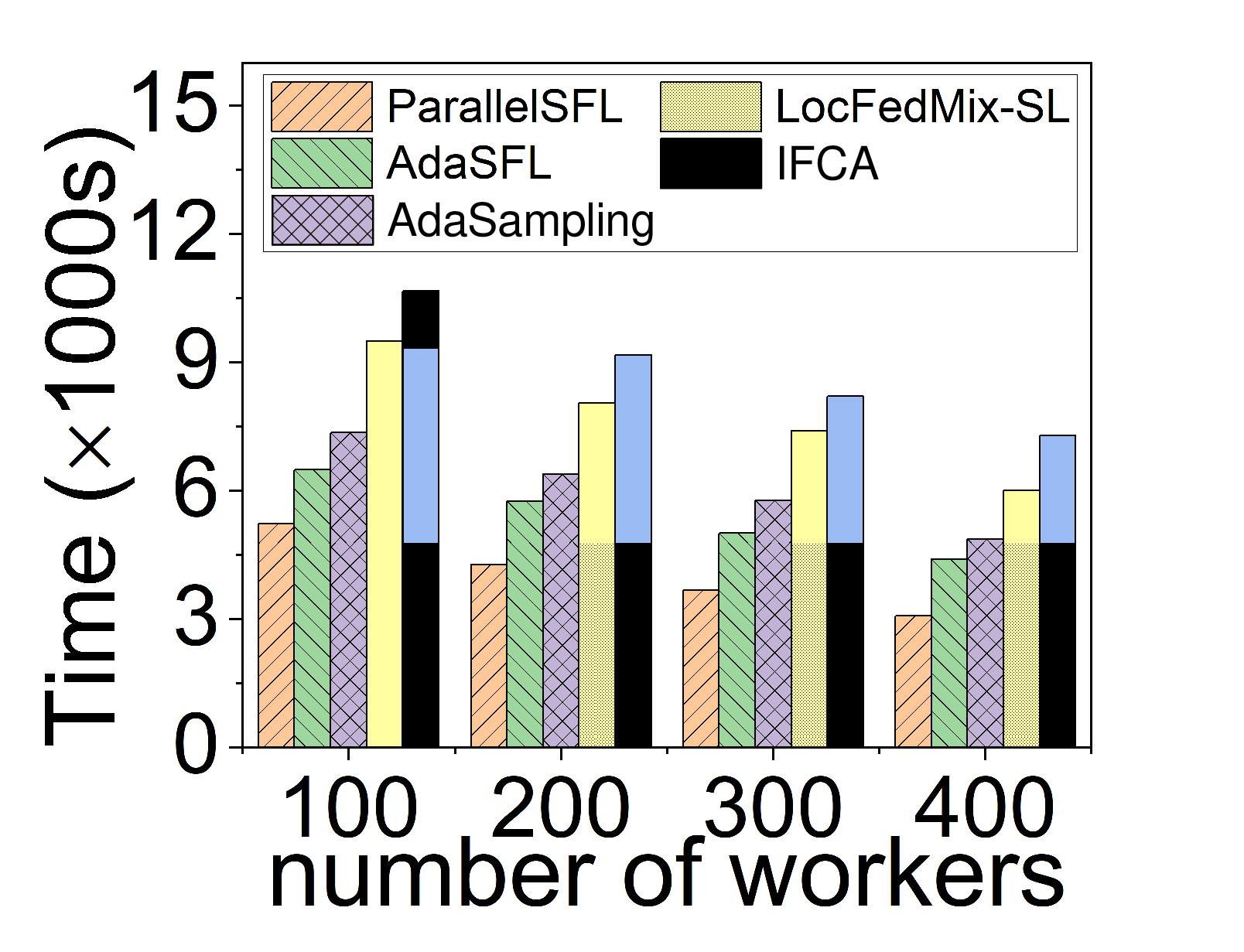}
    \label{fig:scales_time}
}\quad
\subfigure[Training Process]
{
    \includegraphics[width=0.45\linewidth,height=3.3cm]{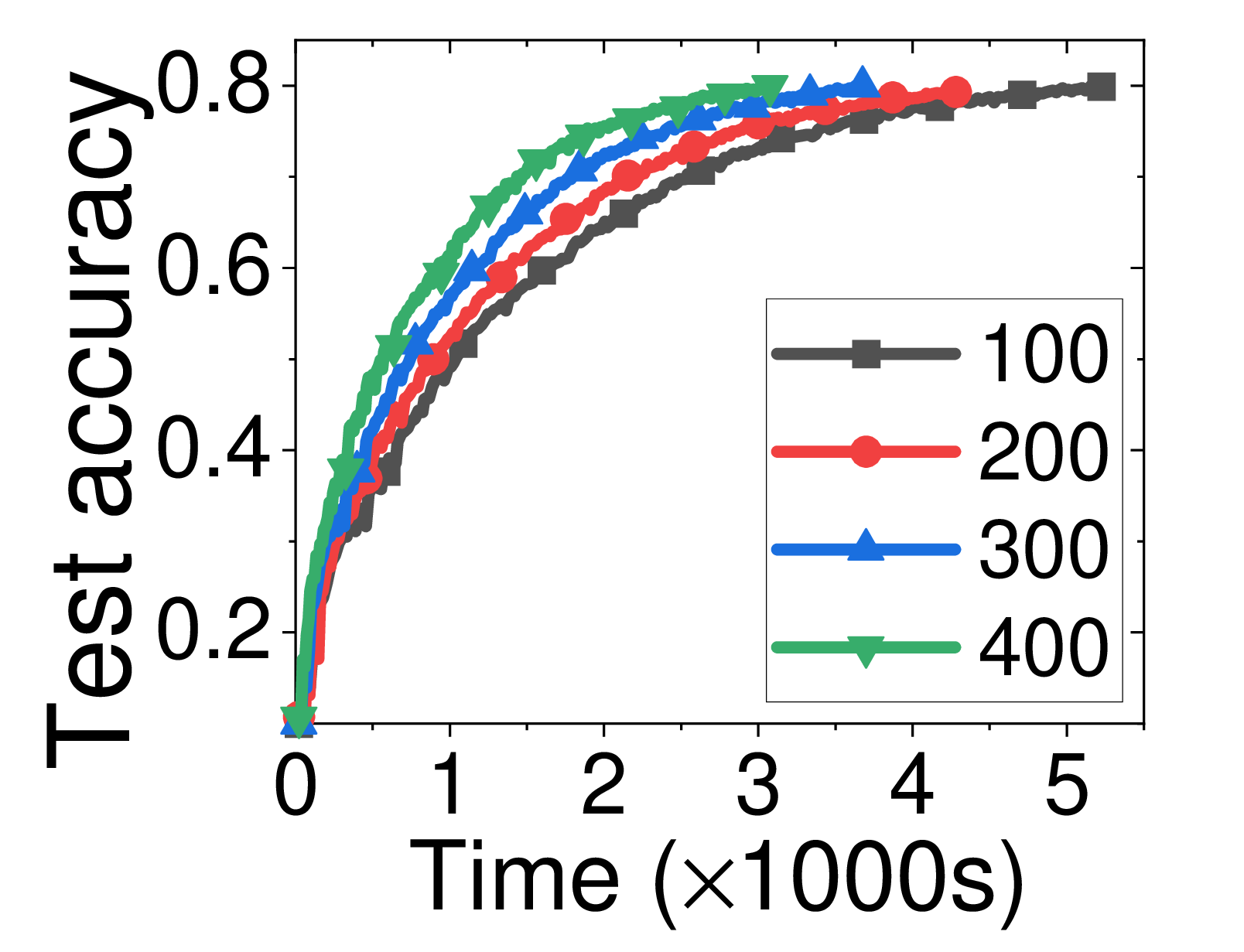}
    \label{fig:scales_time_acc}
}
\caption{Performance comparison with different number of workers.}
\label{fig:scales}
\end{figure}

\subsection{Effect of Non-IID Levels}
To demonstrate the effectiveness of \method in handling non-IID data, we present the test accuracy of different approaches at varying non-IID levels in Fig. \ref{fig:noniid_level}, in which the model accuracy of the five approaches on all the datasets decreases as the non-IID level increases.
However, \method consistently outperforms the other approaches on all datasets.
\bluenote{
AdaSFL and LocFedMix-SL, without considering the challenges of statistical heterogeneity, exhibit the lowest model accuracy on non-IID datasets.
To tackle statistical heterogeneity, IFCA, which estimates the cluster identities, and AdaSampling, which designs an adaptive worker sampling algorithm, mitigate the impact of non-IID data on model training to some extent.
Specifically, as illustrated in Fig. \ref{fig:noniid_level_speech}, \method can achieve improvement of test accuracy by about 19.61\%, 2.54\%, 20.33\%, and 15.13\% on Speech with non-IID level of $p$=10, compared to the baselines (\ie, AdaSFL, AdaSampling, LocFedMix-SL and IFCA).
Notably, by Fig. \ref{fig:noniid_level_CIFAR10}, with non-IID level of $p$=10 on CIFAR-10, \method achieves an improvement of final accuracy by about 21.17\%, 9.65\%, 22.15\%, 16.84\%, compared to the baselines (\ie, AdaSFL, AdaSampling, LocFedMix-SL, IFCA).
Besides, as shown in Fig. \ref{fig:noniid_level_image}, while transitioning from IID to non-IID level of $p$=10 on IMAGE-100, \method, AdaSampling and IFCA suffer from only 13.35\%, 17.41\% and 21.64\% loss in accuracy, while the accuracy loss for AdaSFL, and LocFedMix-SL is 24.89\% and 24.86\%, respectively.
Moreover, by Fig. \ref{fig:noniid_level_Bert}, with non-IID level of $p$=10 on QNLI, \method, AdaSampling and IFCA achieve 90.11\%, 87.33\%, and 85.76\% accuracy, while LocFedMix-SL and AdaSFL only achieve 84.51\% and 84.62\% accuracy.
Collectively, these results further demonstrate the advantage of \method with effective cluster clustering in addressing statistical heterogeneity.
}

\subsection{Effect of Key Strategies}
There are two key strategies of \method, \ie, worker clustering and local updating frequency optimization, being developed to enhance the performance of SFL.
Herein, we conduct several sets of experiments for training AlexNet on CIFAR-10 with IID distribution ($p$=0) and non-IID distribution ($p$=10) to evaluate the effectiveness of the two critical strategies.
We adopt the \method without local updating frequency optimization (\method-WC) and typical SFL with random worker clustering (\method-R) as the baselines.
Concretely, in \method-WC, the PS assigns the identical and fixed local updating frequency for each cluster, while in \method-R, the PS randomly partitions the workers into several clusters and determines the top worker for each cluster with identical and fixed local updating frequency.
By Fig. \ref{fig:key_strategies}, \method-WC converges much faster than \method-R on the IID dataset, and \method-WC also achieves higher test accuracy than \method-R on the non-IID dataset.
Specifically, \method-WC reduces the total training time by about 38\% and improves the final test accuracy by about 20.06\% on the non-IID dataset, compared to \method-R.
The results illustrate that our worker clustering strategy is effective in addressing the system and statistical heterogeneity.
Besides, powered by the local updating frequency optimization among worker clusters, \method speeds up training by about 1.56$\times$ compared to \method-WC on both IID and non-IID datasets.
The results reflect the positive roles of worker clustering and local updating frequency optimization in \method.

\subsection{Effect of System Scales}
In this section, to demonstrate the robustness of \method, we evaluate the performance of \method and baselines with different scales of participating workers.
We conduct several sets of experiments for training AlexNet on CIFAR-10 with four scales (\eg, 100, 200, 300, 400) through extensive simulation experiments, which are conducted on an AMAX deep learning workstation equipped with an Intel(R) Xeon(R) Platinum 8358P CPU @ 2.60GHz, 4 NVIDIA GeForce RTX A6000 GPUs (48GB GPU memory each) and 512 GB RAM.
The results of completion time to achieve 80\% accuracy for these approaches are presented in Fig. \ref{fig:scales_time}, while the training processes of different scales for \method are presented in Fig. \ref{fig:scales_time_acc}.
As the number of participating workers increases, all approaches achieve faster convergence.
The reason is that the number of samples on a worker is limited and more workers contribute more local data for model training in each round, thus speeding up model training.
For instance, \method with 400 workers reduces the total training time by about 41\%, 28\%, 16\%, compared to \method with 100, 200, and 300 workers, respectively.
\bluenote{In addition, \method also achieves a speedup of 1.33$\times$$\sim$2.27$\times$ to reach the target accuracy, compared to the baselines (\ie, AdaSFL, AdaSampling, LocFedMix-SL, IFCA) regarding the different scales of participating workers.}
These results further illustrate the robustness and advantage of \method.

%% file: contents/conclusion.tex
In this paper, we have proposed a novel SFL framework, named \method, which integrates the advantages of FL and SFL and is designed to tackle heterogeneity issues by effective cluster partitioning.
Under the heterogeneity restrictions, \method defines a utility function to serve as a comprehensive metric for estimating waiting time and data distribution of worker clusters, and it contributes to balancing the trade-off between training efficiency and model accuracy.
The experimental results show that the \method can speed up the model training by at least 1.36$\times$ and improve model accuracy by at least 5\% in heterogeneous scenarios, compared to the baselines.